\documentclass[useAMS,usenatbib]{mnras}
\usepackage{amsmath}
\usepackage{amssymb}
\usepackage{graphicx}

\newcommand{\ESN}{E_{\rm SN}}
\newcommand{\CFe}{\mbox{[C/Fe]}}
\newcommand{\FeH}{\mbox{[Fe/H]}}

\newcommand{\lturb}{\ell_{\rm turb}}
\newcommand{\vturb}{v_{\rm turb}}

\newcommand{\MACH}{M_{\rm ACH}}
\newcommand{\Mmix}{M_{\rm mix}}
\newcommand{\Msun}{\rm{M}_\odot}

\newcommand{\zvir}{z_{\rm vir}}
\newcommand{\nflat}{n_{\rm flat}}

\newcommand{\RLACH}{R_{\rm L,ACH}}

\title[Preserving primordial chemical signatures]{Preserving
  chemical signatures of primordial star formation in the first low-mass stars}
\author[Alex P. Ji, A. Frebel, V. Bromm]{Alexander P. Ji$^{1}$\thanks{E-mail:
alexji@mit.edu}, Anna Frebel$^{1}$ and Volker Bromm$^{2}$\\
$^{1}$Kavli Institute for Astrophysics and Space Research and
Department of Physics, Massachusetts Institute of Technology, \\ 77
Massachusetts Avenue, Cambridge, MA 02139, USA\\
$^{2}$Department of Astronomy, University
of Texas at Austin, 2511 Speedway, Austin, TX 78712, USA}
\begin{document}

\date{Submitted July 2015}

\pagerange{\pageref{firstpage}--\pageref{lastpage}} \pubyear{2015}

\maketitle

\label{firstpage}

\begin{abstract}
We model early star forming regions and their chemical enrichment by
Population~III (Pop~III) supernovae with nucleosynthetic yields
featuring high [C/Fe] ratios and pair-instability supernova (PISN) signatures.
We aim to test how well these chemical abundance signatures are
preserved in the gas prior to forming the first long-lived low-mass
stars (or second-generation stars). 
Our results show that second-generation stars can retain the
nucleosynthetic signature of their Pop~III progenitors, even in the
presence of nucleosynthetically normal Pop~III core-collapse supernovae.
We find that carbon-enhanced metal-poor stars are likely
second-generation stars that form in minihaloes. Furthermore, it is
likely that the majority of Pop~III supernovae produce high [C/Fe]
yields.
In contrast, metals ejected by a PISN are not concentrated in the
first star forming haloes, which may explain the absence of observed
PISN signatures in metal-poor stars.
We also find that unique Pop~III abundance signatures in the gas are
quickly wiped out by the emergence of Pop~II supernovae. 
We caution that the observed fractions of stars with Pop~III signatures
cannot be directly interpreted as the fraction of Pop~III stars
producing that signature. Such interpretations require modelling
the metal enrichment process prior to the second-generation stars'
formation, including results from simulations of metal mixing.
The full potential of stellar archaeology can likely be reached in
ultra-faint dwarf galaxies, where the simple formation history may
allow for straightforward identification of second-generation stars.
\end{abstract}


\section{Introduction} \label{sec:intro}
The formation of the first, so-called Population~III (Pop~III), stars
marks an important yet incompletely understood period in
cosmic history. As the first non-linear baryonic structures in the
Universe, Pop~III stars initiate the reionization and chemical
enrichment of the early intergalactic medium, thus setting the stage
for the assembly of more mature galaxies (e.g. \citealt{Bromm09};
\citealt{LoebF13}, and references therein).

The character of Pop~III star formation has been theoretically
investigated in great detail, and simulations of this process have
begun to converge on consistent conclusions \citep{Bromm13,Glover13,Greif15}.
Unfortunately, observations that can directly constrain these ideas are in short
supply. The James Webb Space Telescope (JWST)
may be able to observe supernovae (SNe) from the most massive Pop~III
stars \citep{Hummel12,Pan12,Whalen13,Chen14,deSouza14}.
The lack of observed low-mass Pop~III stars in the Milky Way can
potentially constrain the low end of the Pop~III initial mass function (IMF),
a constraint that will gain in strength with increasing survey sizes \citep{Hartwig15}.
The ionizing effect of Pop~III stars may be seen in epoch of reionization
measurements \citep{Greif06}, although this prospect has significantly
weakened in light of the recent {\it Planck} determination of a lower
optical depth to Thomson scattering (\citealt{Planck15}).
Arguably, in the absence of such direct probes, the best available constraints on
Pop~III star formation may come from chemical abundances in the most
metal-poor stars, which preserve the chemical composition of their birth
clouds in their stellar atmospheres (e.g. \citealt{Karlsson13, Frebel15}). 
The ideal objects for probing Pop~III star formation are the first
metal-enriched low-mass stars (or \emph{second-generation stars}), which
contain only metals from Pop~III stars.

It is not known with any certainty what abundance signatures would
clearly indicate enrichment primarily by Pop~III stars, but intriguing
empirical hints have been found.
Some unique abundance patterns have been identified in metal-poor
stars, including the high fraction of carbon-enhanced metal-poor
(CEMP) stars at extremely low iron abundance (e.g. \citealt{Norris13,Placco14}), and
a strong odd-even effect characteristic of pair-instability supernovae (PISNe)
\citep{Heger02,Aoki14}.
However, average metal-poor halo stellar abundances ($\mbox{[Fe/H]}<-2.5$)
suggest that the majority of Pop~III stars died as core-collapse
supernovae (CCSNe) \citep{Joggerst10}.
As the ejecta from multiple SNe mix together, any unique Pop~III
abundance patterns which could be preserved in metal-poor stars will
be diluted away to an average core collapse SN yield.

Recent simulations have suggested that
second-generation stars should generically form by combining the metal
yields from more than one Pop~III SN explosion, even in the simplest sites of second
generation star formation \citep{Ritter15,Jeon15}.
Such an immediate enrichment by multiple SNe could rapidly
dilute away unique Pop~III chemical signatures in the majority of the
subsequent, Pop~II, star formation events.
Modeling such dilution from multiple SNe is a classical problem in the theory
of chemical evolution. Several approaches have been employed, including
models that assume instantaneous mixing (e.g.
\citealt{Tinsley80,Kirby11}), extensions that allow inhomogenous
mixing (e.g. \citealt{Oey00, Argast04, Karlsson05}), and full
hydrodynamic simulations (e.g. \citealt{Kobayashi11,Webster14}). 
These models were all devised to study chemical evolution in a static
self-enriching system over multiple generations, such as the Milky Way
disc or a large dwarf galaxy, with the exception of
\citealt{Webster14} who specifically investigate chemical evolution in
ultra-faint dwarf galaxies.
However, second-generation stars form in a unique initial stage of
chemical enrichment, taking place in the earliest, low-mass building
blocks of galaxy formation. The existing chemical evolution models
focus less on the initial chemical enrichment stage and more on the
subsequent chemical evolution.

In this paper, we investigate how mixing the ejecta from multiple SNe
may remove tell-tale nucleosynthetic signatures from Pop~III stars.
We consider second-generation star formation in the simplest possible
environments motivated by cosmological simulations of the early Universe.
Considering these environments, we apply idealized mixing models 
to predict when the Pop~III signature is preserved by metal-poor stars
and can be probed by stellar archaeology.

The structure of our paper is as follows.
In Section~\ref{sec:env}, we discuss what is known about where and how stars form
in the early Universe. 
In Section~\ref{sec:mix}, we describe mixing models to produce chemical
distributions both for second-generation stars and subsequent
generations of stars. 
The results of our models are presented in Section~\ref{sec:results}, and we conclude 
in Section~\ref{sec:conclusion}.
Whenever needed, we use a cosmology with parameters $\Omega_m = 0.3$,
$\Omega_\Lambda = 0.7$, $h = 0.7$, and $\Omega_b = 0.05$. The baryon
fraction then is $f_b = 1/6$.
When gas is ionized and neutral, we use $\mu_{\rm ionized} = 0.59$ and
$\mu_{\rm neutral} = 1.23$, respectively, where $\mu m_p$ is the mean
molecular weight of the gas.

\section{Star Formation in the Early Universe} \label{sec:env}
\begin{figure}
  \includegraphics[width=8cm]{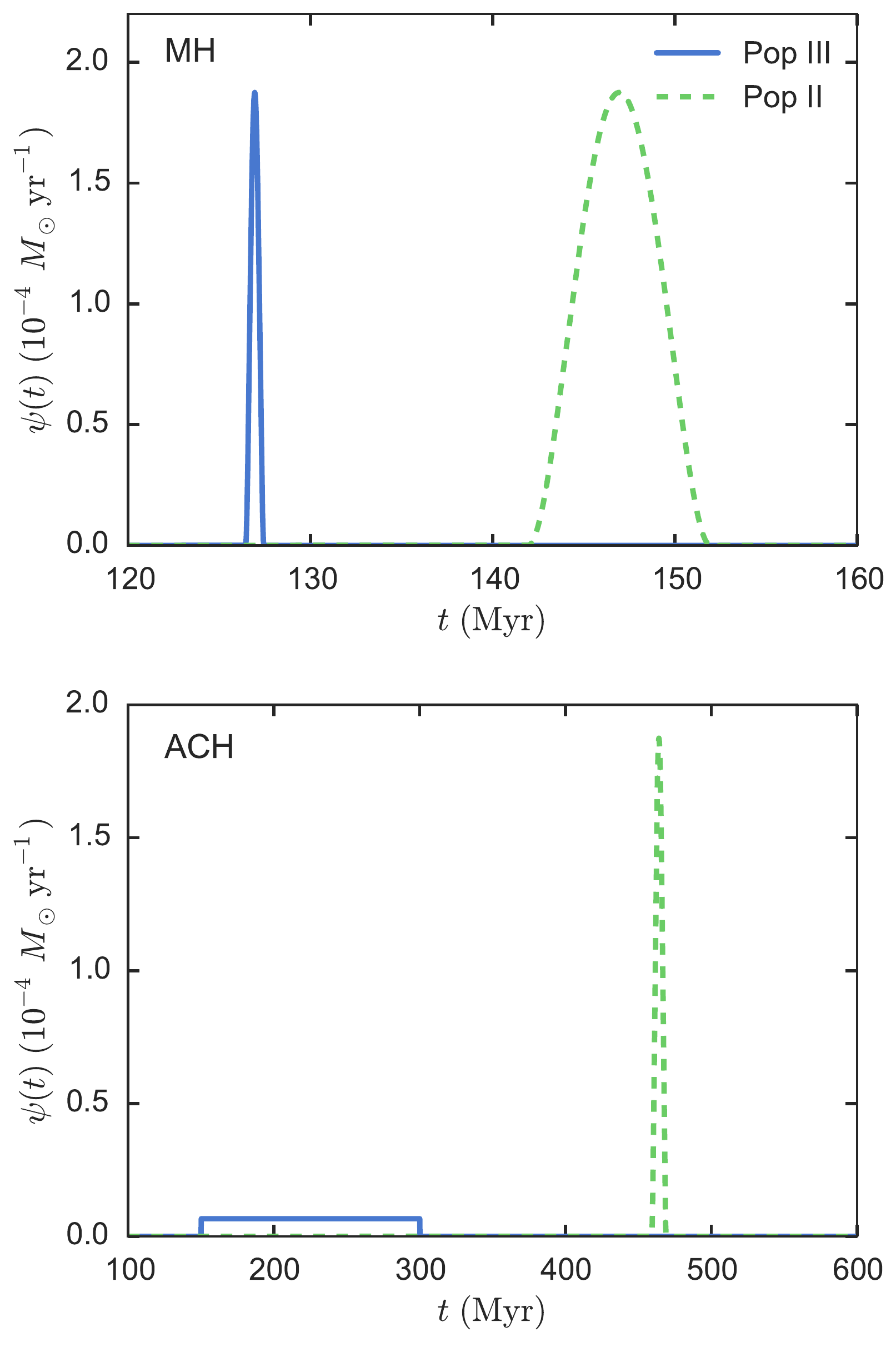}
  \caption{Schematic early star formation rates.
    Top: minihalo star formation rates. Pop~III forms $100
    \Msun$ in a $10^5$ yr burst at virialization ($z=25$). Pop~II
    forms $1000 \Msun$ in a $10$\,Myr burst after a recovery time of
    $20$\,Myr.
    Bottom: atomic cooling halo star formation rates. Pop~III forms
    $1000 \Msun$ uniformly over $150$\,Myr following the merger tree
    in \citet{Greif08}. A Pop~II burst at virialization ($z=10$) forms
    $1000 \Msun$ of stars over $10$\,Myr.
    \label{fig:sfr1}}
\end{figure}

 To model the first chemical enrichment events, we must consider the nature 
of star formation in the early Universe. Here, we take an ab initio approach to 
understanding early star formation based on theoretical arguments and
review the insights obtained from simulations. The
picture described here motivates the mixing models presented in Section~\ref{sec:mix}.

Structure formation in $\Lambda$CDM is a hierarchical process, with
small dark matter haloes forming early and merging into larger haloes.
Two environments stand out in the early Universe where first star formation
can take place. These environments
correspond to the two most significant cooling mechanisms in metal-poor gas. The first stars 
in the Universe form in minihaloes, where star formation is triggered by molecular 
hydrogen cooling (e.g. \citealt{Tegmark97}). The first galaxies in the Universe 
likely form in atomic cooling haloes, when hydrogen line cooling becomes 
significant (e.g. \citealt{Oh02}).
Cosmological simulations of these two environments have outlined a general 
picture for the formation of Pop~III stars and the first few generations of 
Pop~II stars, which we describe in Sections~\ref{envsec:III} and \ref{envsec:II}. 
The star formation rates described are shown schematically in Figure~\ref{fig:sfr1}. 
We discuss nucleosynthetic signatures from the first generations of stars in Section~\ref{envsec:yields}.

\begin{figure}
  \includegraphics[width=8cm]{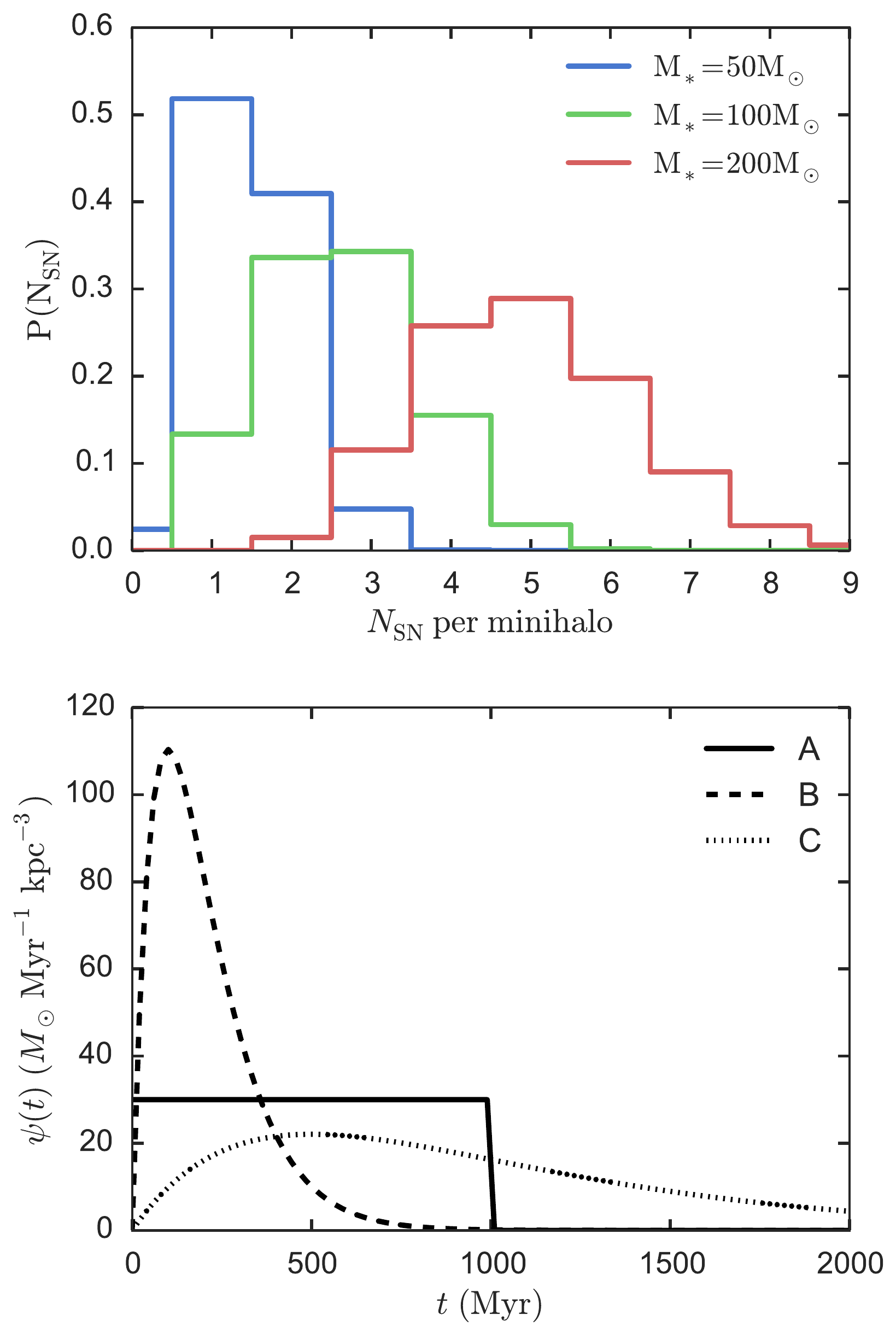}
  \caption{
    Top: distribution of number of Pop~III SNe per minihalo,
    assuming $100 \Msun$ of Pop~III stars form. This was computed
    as described in the text from $10^6$ minihaloes, assuming stars
    with $M \geq 10\Msun$ turn into SNe.
    Bottom: Pop~II star formation rates for a self-enriching
    atomic cooling halo. Model A is a 1\,Gyr flat star formation
    rate. Model B is a $100$\,Myr burst. Model C is a $500$\,Myr
    burst.
    \label{fig:sfr2}}
\end{figure}

\subsection{Population III star formation} \label{envsec:III}
\subsubsection{Pop~III initial mass function}
Pop~III stars form in minihaloes of mass $\sim 10^6\Msun$ around
$z \sim 25$, where star formation is triggered by molecular hydrogen cooling
\citep{Couchman86,Haiman96,Tegmark97}.
The lack of metal cooling and the resulting higher temperatures
suggest that Pop~III stars have high characteristic masses, but
our understanding of the characteristic mass has changed
considerably over the last decade (e.g. \citealt{Bromm13}).
Early simulations found that these stars might have characteristic
masses $\gtrsim 100 \Msun$ with one star forming per
minihalo \citep{Abel02,Bromm02,Yoshida06,OShea07}. These stars would
typically explode as PISNe with explosion energies $\gtrsim 10^{52}$\,erg (e.g. \citealt{Heger02}).
The strong feedback from these stars likely prevents further star formation within
their host minihaloes \citep{Greif07,Whalen08}.

However,
more recent studies with higher spatial resolution have found that
fragmentation may occur in the disc surrounding a Pop~III protostar,
forming small multiples of stars with lower characteristic masses
$\gtrsim 10 \Msun$ \citep{Clark08,Clark11,Stacy10,Greif11,Greif12}.
The amount of fragmentation varies between different minihaloes and
depends on the detailed thermal history of the gas \citep{Greif12}.
This general picture has been corroborated by observations of 
metal-poor stars, which seem to show clear $\alpha$-element
enhancements that are consistent with core-collapse supernovae
(e.g. \citealt{Beers05}).

Although a shape and range of the IMF cannot yet be determined robustly,
\citet{Greif11} find that the protostellar mass function across five
minihaloes is approximately logarithmically flat ranging from $\sim
0.1-10 \Msun$ based on 1000 years of fragmentation and accretion,
although accretion should continue for $10^5$ or $10^6$ years in total.
\citet{Susa14} evolve 59 minihaloes with lower resolution for $\sim
10^5$ years after sink particle formation, also finding a fairly
logarithmically flat IMF ranging from $\sim 1-100 \Msun$. 
Such simulations suggest that the Pop~III IMF is close to
logarithmically flat from $\sim 1-100 \Msun$ (with a characteristic
mass $\sim 20 \Msun$), though a flat IMF is also reasonable (e.g.
\citealt{Cooke14})\footnote{A logarithmically flat IMF has $dN/d\log M\propto \text{constant}$, 
  while a flat IMF has $dN/dM \propto \text{constant}$.}.
With such an IMF, it is likely that multiple CCSNe form within each minihalo.
Thus, we use a log-flat IMF with lower and upper mass bounds of $1$
and $100 \Msun$.

The mass range suggested here does not allow for higher mass
PISNe. If PISNe occur, the PISN progenitor is likely the
only Pop~III star to form in a given minihalo.
There is currently only tentative evidence for metal-poor stars with
PISN abundance patterns \citep{Aoki14}, because
PISN signatures may only be found in higher metallicity stars and thus
be missed by current metal-poor star surveys which preferentially
select the most metal-poor stars \citep{Karlsson08}. 
In addition, superluminous supernovae which might be associated with
PISNe have been observed at both high and low redshift
\citep{GalYam09,Cooke12}. If a PISN explodes in a minihalo, the
violent feedback associated with it will likely prevent any subsequent
star formation there.

\subsubsection{Pop~III star formation rate}
The Pop~III star formation in an individual minihalo should occur
in one burst. Collapse begins around the virialization time of the
dark matter halo, and stars form in about a free fall time.
The mass accretion rates into the central star forming regions are
around $10^{-3}-10^{-2} \Msun\, \rm{yr}^{-1}$ for $\sim 10^5$ yr (e.g.
\citealt{Greif11,Hirano14,Susa14}).
Not all of this mass ends up in protostars due to radiation feedback
\citep{Stacy12}, and each minihalo ends up forming around $100
\Msun$ of stars (e.g. \citealt{Susa14}).

Although our Pop~III IMF allows low mass stars to form, for the
investigation of metal enrichment we are only
interested in high mass stars which explode as SNe. 
The exact number of SNe in a minihalo is not well determined;
simulations which resolve significant fragmentation do not run long
enough to determine the actual stellar masses (e.g.
\citealt{Greif11}), while simulations running for longer times may not
fully resolve the fragmentatien (e.g. \citealt{Hirano14,Susa14}).
As a simple way
to model the number of Pop~III SNe exploding within a minihalo,
we sequentially draw stellar masses from the log-flat IMF ranging over
$1-100 \Msun$ until we reach a total of 
$\rm{M}_* = 50,\,100,\,\rm{or\ }200\Msun$ of stars. If adding a star
would cause us to increase the mass beyond $\rm{M}_*$, we 
redraw from the IMF. Only stars with mass larger than $10 \Msun$ are
assumed to undergo a supernova, so we stop when we have drawn 
$>\rm{M}_*-10\Msun$ of stars.
The distribution of the number of SNe found this way from $10^6$
minihaloes is shown in Figure~\ref{fig:sfr2}.

This procedure gives initial stellar masses in addition to the total
number of SNe in a minihalo, and in principle the stellar masses can
be used as inputs to SNe nucleosynthesis calculations to determine the
metal yields on a star-by-star basis.
However, given the uncertainties in these calculations (see
Section~\ref{envsec:yields}), we aim to separate the impact of SN yields
and energies from the problem of metal mixing.
Thus, we do not use the stellar masses derived from this procedure for
anything other than estimating the number of SNe in a minihalo.

Pop~III stars will not form in atomic cooling haloes, unless in rare
cases, because the haloes will typically be polluted with metals prior
to their virialization (e.g. \citealt{Johnson08}).
One can instead estimate the number of star forming
minihaloes that eventually merge into an atomic cooling halo.
\citet{Greif08} followed one atomic cooling halo and found that the 
ejecta from $\sim 10$ star forming minihaloes mixed together during the 
formation of the atomic cooling halo, with the Pop~III star formation 
occurring from $t=150-300$\,Myr. If each minihalo produces $\sim
100 \Msun$ of Pop~III stars, the Pop~III star formation rate is
about $1000 \Msun$ in $150$\,Myr.
For an extreme upper bound, a maximum of one hundred $10^6 \Msun$ minihaloes 
can make up a single $10^8 \Msun$ atomic cooling halo.

\subsection{Population II star formation} \label{envsec:II}
In this section we describe the formation of Pop~II stars. The transition
from Pop~III to Pop~II is often demarcated by a critical metallicity.
The critical metallicity could be set by atomic line cooling 
(e.g. \citealt{Bromm03,Frebel07})
or dust thermal cooling (e.g. \citealt{Schneider12,Ji14}).
For simplicity, we do not apply any critical metallicity, and instead 
assume all metal-enriched gas is able to form low mass Pop~II stars 
that could correspond to metal-poor stars. 
In practice, our modelled metallicities are nearly always above 
$Z \sim 10^{-5} Z_\odot$, which places all gas above the critical 
metallicity for low-mass star formation according to the dust 
cooling criterion.

\begin{table}
  \caption{Second-generation star forming environments \label{tbl:env}}
  \begin{tabular}{lcc}
    \hline
     & Minihalo & Atomic Cooling Halo \\
    \hline
    Halo Mass & $10^6 \Msun$ & $10^8 \Msun$ \\
    $z_{\rm vir}$ & 25 & 10 \\
    $R_{\rm vir}$ & 120 pc & $1.3$\,kpc \\
    $v_{\rm vir}$ & 6 km s$^{-1}$ & 18 km s$^{-1}$\\
    $T_{\rm vir}$ & 1900 K & 17400 K \\
    $D_t$$^a$ [kpc$^2$\,Myr$^{-1}$] & $2.4 \times 10^{-5}$ & $8.1 \times 10^{-4}$\\
    \hline
  \end{tabular}
  \\
  $^a$ effective diffusion coefficient for turbulent mixing
\end{table}

\subsubsection{Pop~II star formation environment}
Both the minihalo and atomic cooling halo environments may host 
early Pop~II star formation. Which environment is relevant in a given
region of space depends on the degree of Pop~III feedback and the
corresponding recovery time \citep{Jeon14}.

If Pop~III feedback is strong, then the minihalo loses most
of its gas (e.g. \citealt{Whalen08}) and the next time gas can collapse
is in atomic cooling haloes with virial temperature $\sim 10^4$\,K or 
mass $\sim 10^8 \Msun$ \citep{Oh02,Greif08,Wise12a}. 
These haloes are perhaps the best candidates for sites of 
first galaxies \citep{Bromm11}.
If instead Pop~III feedback is weak, then it is possible for gas to
recollapse in the original minihalo and form Pop~II stars there
(e.g. \citealt{Ritter12,Ritter15,Cooke14}). These
stars form out of a mix between the supernova ejecta that have fallen
back and primordial gas inflowing to the centre of the minihalo
through filaments. 
The minihalo may gain some mass during the recollapse \citep{Jeon14}.

It is likely that differing accretion/merger histories and the
distribution of Pop~III stellar masses result in different recovery
times, allowing second-generation stars to form from both of these 
environments. The atomic cooling halo and minihalo will be the two
main contributers to the population of second-generation stars in the
$\Lambda$CDM paradigm. However, there are two other situations that
may arise.
It is possible that the blast from a PISN in one minihalo is able
to pollute a nearby minihalo and trigger second-generation star
formation (e.g. \citealt{Greif10,Wise12a,Smith15}), although the
degree of metal pollution may not be very high \citep{Cen08}.
Also, star formation may be triggered in the shocked shell of a 
Pop~III supernova explosion (e.g. \citealt{Mackey03, Chiaki13b}). 
These two cases are possible variants on the picture of metal recollapse
in a minihalo which depend on the specific hydrodynamics of metal
mixing and supernova shocks, and we do not expect the majority of
second-generation stars to form through such mechanisms. However, such
cases may be able to sample the metal output of individual SNe.

As canonical values, we consider minihaloes of mass $\sim 10^6 \Msun$
virializing at $z\sim 25$, and in case of moderate
stellar feedback we assume second-generation (Pop~II) star formation
takes place in a similar environment. For the atomic cooling halo
case, we assume a typical halo mass of $\sim 10^8 \Msun$,
virializing at $z\sim 10$ \citep{Greif08}.
Note that in both these environments, there will be a
distribution of halo masses and virialization redshifts
\citep{Wise12a,Hirano14,Susa14}. The halo mass is especially important
in the minihalo case as a higher halo mass increases the likelihood
that second-generation stars can form within the minihalo \citep{Cooke14,Jeon14}.
Canonical physical parameters of these haloes are given in Table~\ref{tbl:env}.

Regardless of whether a second-generation star forms in a minihalo or
an atomic cooling halo, it is very likely that metals from
multiple Pop~III SN will contribute to the second-generation star
forming gas, for the following reasons. If the second-generation star
forms in an atomic cooling halo, its metal content is determined by
mixing together metals from many progenitor minihaloes. 
If instead the second-generation star forms in a
minihalo, the Pop~III stars which formed in that minihalo were of lower
mass, resulting in less vigorous feedback. Thus, it is likely that
a small multiple of Pop~III stars contributed metals to
second-generation stars.

\subsubsection{Pop~II star formation rates}
When Pop~II star formation occurs in a minihalo, the SN ejecta
are mixed with pristine infalling gas, accreting through filaments,
in a turbulent core, collapsing after a recovery time of
$\sim 20$\,Myr \citep{Jeon14,Ritter15}. 
The rate of Pop~II star formation in this case has not been simulated,
but in the study by \citet{Ritter12} the cool dense gas core at
the centre has a mass $\sim 3000 \Msun$. If the star formation
efficiency is $\sim 30$ per cent, then this will ultimately produce $\sim
1000 \Msun$ of stars in $\sim 10$\,Myr. 
After $10$\,Myr, the more massive Pop~II stars will explode as
supernovae ($\sim 5$SNe).
If a sufficient number of SNe explode and the minihalo is not too
massive, this will finish the gas expulsion initiated by the Pop~III
stars and halt star formation in the minihalo until it merges into
another system. On the other hand, if the minihalo has gained enough
mass to contain the SN explosion, then the minihalo
can continue to self-enrich and form stars out of gas contaminated by
Pop~II SNe \citep{Jeon14}.

Pop~II star formation in an atomic cooling halo begins approximately
when the halo virializes ($z\sim 10$ or $t\sim 400$\,Myr) and atomic
line cooling triggers star formation (e.g. \citealt{Greif08, SafShrad14b, SafShrad14}).
The first burst of star formation will occur near the centre of the
atomic cooling halo, after which star formation will be temporarily
halted. Stars forming after this initial $\sim 10$\,Myr burst of Pop~II star
formation will no longer be truly second-generation stars, having been
polluted by the first generation of Pop~II stars.
The amount of stars formed in this burst is not certain. 
\citet{Greif08} find that $10^5 \Msun$ of cold dense gas has
collapsed with $Z \sim 10^{-3} Z_\odot$. 
\citet{SafShrad14} follow collapse to a gas density of
$10^7$\,cm$^{-3}$ and find that stellar clusters can form with $\sim
1000 \Msun$ of stars after $4$\,Myr. 
\citet{SafShrad15} resolve the same initial conditions down to 
a density of $\sim10^{13}$\,cm$^{-3}$, resolving individual
protostellar cores as sink particles and including the effect of their
radiative feedback. In 20,000\,yr, $\sim 80 \Msun$ of stars formed, although
many of the protostars will continue to accrete gas beyond the
simulated time.
Thus it appears that approximately $1000 \Msun$ of second-generation
stars can form in this initial $10$\,Myr burst.
The SNe from this burst likely contaminate the dense gas with metals
from Pop~II stars \citep{Wada03}, preventing further second-generation stars from
forming afterwards.

After the initial burst, Pop~II star formation in the atomic cooling
halo will continue until it is halted. This could happen externally
through reionization or merging into another halo, or it could happen
internally through SN feedback.
We estimate these processes with three models. Model A is a constant
star formation rate for 1\,Gyr followed by an abrupt cutoff. Model B is
a short $\sim 100$\,Myr burst. Model C is a longer $\sim 500$\,Myr
burst. The burst in models B and C use the equation
\begin{equation} \label{eq:sfrshape}
  \psi(t) = \psi_0 \left(\frac{t}{\tau}\right) e^{-t/\tau}
\end{equation}
where $\psi_0$ is some normalization and $\tau$ is the characteristic
time of the burst ($100$\,Myr and $500$\,Myr for models B and C, respectively).
These star formation rates are plotted in Figure~\ref{fig:sfr2}. The
normalizations are all set to produce 30,000\,$\Msun$\,kpc$^{-3}$.

\subsubsection{Pop~II initial mass function}
Modern star formation theory has recognized the importance of
supersonic turbulence in regulating star formation \citep{MacLow04,McKee07}.
The turbulent velocity helps prevent global gravitational collapse,
while the complex network of shocks formed by supersonic flows create
local density fluctuations that can trigger star formation.
The turbulent density spectrum may play a fundamental
role in determining the IMF \citep{Hopkins12}.

In atomic cooling haloes, supersonic turbulence is created during
virialization of the halo, and during subsequent accretion from the
cosmic web (e.g. \citealt{Greif08, SafShrad14, SafShrad14b}). 
This suggests that the Pop~II IMF in atomic cooling
haloes may be similar to that in present-day star formation. We thus take
the IMF in atomic cooling haloes to be the Salpeter IMF from
$0.1-100\,\Msun$. 
To date, there have not been significant studies of turbulence in
minihaloes which have experienced recollapsing gas. We assume the
IMF in this situation is also Salpeter.

\subsection{Nucleosynthetic signatures} \label{envsec:yields}
\begin{table}
  \caption{CEMP signature supernova yields \label{tbl:CEMPyields}}
  \begin{tabular}{lcccc}
    \hline
    & CCSN$^{a}$ & Faint SN$^{b}$ & Wind+SN$^{c}$ & Pop~II SN$^{d}$\\
    \hline
    Mass ($\Msun$) & 20 & 20 & 20 & 20 \\
    $E_{\rm SN,51}$$^{e}$  & $1$ & $0.74$ & $1$ & $1$ \\
    C ($\Msun$)    & 0.211   & 0.20                & 1.034 & 0.128 \\
    Fe ($\Msun$)      & 0.072   & 1.09 $\times 10^{-5}$ & 0.072 & 0.073 \\
    \hline
  \end{tabular}
\raggedright
$^{a}$ \citealt{Nomoto06} \\
$^{b}$ \citealt{Iwamoto05} \\
$^{c}$ CCSN with added C from \citealt{Hirschi07} (see text) \\
$^{d}$ \citealt{Nomoto13} \\
$^{e}$ $E_{\rm SN,51}$ is the explosion energy in units of $10^{51}$\,erg
\end{table}
\begin{table}
  \caption{PISN signature supernova yields \label{tbl:PISNyields}}
  \begin{tabular}{lccc}
    \hline
    & CCSN$^{a}$ & PISN$^{b}$ & Pop~II SN$^{c}$\\
    \hline
    Prog. Mass ($\Msun$)   & 20 & 195 & 20 \\
    $E_{\rm SN,51}$  & $1$ & $40$ & $1$ \\
    C ($\Msun$)    & 0.211   & 4.13 & 0.128 \\
    Na ($\Msun$) & 0.0029& 0.00028 & 0.00181\\
    Mg ($\Msun$) & 0.150   & 4.39 & 0.247 \\
    Ca ($\Msun$) & 0.00623 & 0.993 & 0.00921\\
    Fe ($\Msun$)      & 0.072   & 3.08 & 0.073 \\
    Co ($\Msun$) & 1.5 $\times 10^{-4}$ & 5.59 $\times 10^{-6}$ & $6.21 \times 10^{-5}$\\
    Ni ($\Msun$) & 0.00175& 0.00825 & $7.11 \times 10^{-4}$\\
    \hline
  \end{tabular}
\\
\raggedright
$^{a}$ \citealt{Nomoto06}\\
$^{b}$ \citealt{Heger02}\\
$^{c}$ \citealt{Nomoto13} \\
\end{table}
We now discuss the Pop~III chemical abundance signatures that
may be found in second-generation stars, as well as other early metal
sources.
Unfortunately,
calculating supernovae yields from first principles is a difficult task. Core collapse
supernovae in particular have proven to be difficult to understand
from first principles due to a complex explosion mechanism. After 
the star's core becomes gravitationally unstable and undergoes core
bounce, the outgoing shock stalls and is presumably revived by some
combination of hydrodynamic instabilities and neutrino heating (e.g.
\citealt{Janka12}). 
3D simulations of SN explosions that treat multifrequency
neutrino transport are a recent development, and the results of these
calculations have not yet converged between different
groups (e.g. \citealt{Hanke13,Mezzacappa14}).
Instead of a full hydrodynamic simulation,
CCSN metal yield calculations often parametrize the explosion in
different ways (e.g. \citealt{Kobayashi06,Heger10,Limongi12}). 
These methods agree qualitatively, but not in detail \citep{Nomoto13}.

Rather than looking to reproduce detailed elemental abundances, we
instead focus on the most important broad signatures of 
Pop~III nucleosynthesis.
Many Pop~III CCSN calculations tend to produce abundances
consistent with the stellar halo chemical record (e.g.
\citealt{Tumlinson06,Joggerst10}), suggesting that Pop~III
CCSNe produce a standard $\alpha$-enhanced yield not qualitatively
different from their higher metallicity counterparts. As a representative yield, we
take the $20 \Msun$, zero-metallicity model with a SN explosion
energy of $\ESN=10^{51}$\,erg from \citet{Nomoto06}.

However, there are two abundance signatures that may distinguish
themselves from this typical yield pattern, the carbon enhanced
metal-poor (CEMP) signature and the pair instability supernova (PISN)
signature. We now discuss these signatures in more detail.

The exact yields we use in our future models are given in
Tables~\ref{tbl:CEMPyields} and \ref{tbl:PISNyields}. We have quoted
the yields with a high degree of precision, but the yields will likely
span a range of values. Our adopted values are only meant to
be representative realizations.

\subsubsection{CEMP signature}
If a star has $\mbox{[C/Fe]} > 0.7$, we say it displays the \emph{CEMP
signature} following \citet{Aoki07}\footnote{$\mbox{[X/Y]} = \log_{10}(N_X/N_Y) -
  \log_{10}(N_X/N_Y)_\odot$ for element X,Y}.
It is not known whether the CEMP signature is
created by Pop~III stars, or if there is a non-primordial explanation
related to surface pollution or a critical metallicity (see
\citealt{Norris13}, and \citealt{Nomoto13}, for a comprehensive account
of possibilities, and \citealt{Sluder15} for a mechanism not
previously considered). We proceed under the assumption the CEMP
signature does indeed trace nucleosynthetic yields of Pop~III
stars. There are two classes of mechanisms to create the signature. It
can be produced either by removing iron from or adding carbon to the
standard CCSN yields.

A likely mechanism for removing iron is to allow
some sort of metal fallback during the SN explosion.
This can be accomplished through faint supernovae that undergo
mixing and fallback \citep{Umeda02,Umeda03,Iwamoto05} or a jet-like
explosion \citep{Tominaga07}. 
As a representative fallback yield, we take a faint CCSN with
$\ESN=0.74 \times 10^{51}$\,erg from \citet{Iwamoto05} which reproduces
the detailed abundances of HE~1327$-$2326 \citep{Frebel05}.

A likely mechanism for adding carbon to the yield is by winds from
rotating massive Pop~III stars \citep{Meynet06,Meynet10,Hirschi07}.
To approximate this, we add $0.823 \Msun$ of carbon to our standard
Pop~III CCSN. This represents the carbon winds from a very fast
rotating star of $20 \Msun$ \citep{Hirschi07}, and results in
$\mbox{[C/Fe]}=0.89$ for the wind model.

\subsubsection{PISN signature}
The PISN signature is still a theoretical signature, although it has
been tentatively observed in a metal-poor star \citep{Aoki14}. 
In contrast with CCSNe, the explosion mechanism for PISNe is better
understood and does not need to be put in artificially (e.g.
\citealt{Heger02}). The typical PISN turns a large fraction of its
stellar mass into metals and exhibits suppressed odd-even abundance ratios.
We will consider a star to have the PISN signature if $\mbox{[Co/Ni]}
< -0.5$, where Co and Ni are representative odd and even elements.
We take the yields from \citet{Heger02} for a $195 \Msun$ 
$\ESN \sim 40\times 10^{51}$\,erg explosion as a characteristic PISN yield.

\subsubsection{Other metal sources}
It will be of interest to look at how Pop~II SNe dilute away
the CEMP and PISN abundance signatures.
For Pop~II SN yields, we use the $20 \Msun$, $Z=0.001$ yields from
\citet{Nomoto13} which are similar to the Pop~III CCSNe yields.

There are two other main sources of metals in standard chemical evolution
models: AGB stars and Type Ia SNe.
Low-to-intermediate mass stars undergo an AGB phase, during which they
expel significant amounts of carbon in their winds. A typical
time until an intermediate-mass star undergoes the AGB phase is $\sim
100$\,Myr, much longer than the recovery time if second-generation
stars form in minihaloes, but shorter than the recovery time for an
atomic cooling halo. However, the total metal mass from a single AGB
star's winds is not very large, and its importance in standard chemical
evolution is largely due to the steep slope of the IMF. The top-heavy
Pop~III IMF suggests that AGB stars will not be an important
contributor to elements found in second-generation stars.
Type Ia SNe produce large amounts of iron-peak elements. In principle, there
could be Pop~III binary white dwarfs that produce Type Ias. These also are
unlikely to contribute to second-generation stars, since such
systems may be rare, and the delay time between their formation and
detonation is typically $\sim 1$\,Gyr (although it could be as low as $100$\,Myr,
see \citealt{Maoz12}). For these reasons, we will not consider 
contributions from these two metal sources.

\section{Metal Enrichment Models} \label{sec:mix}
\begin{figure}
  \includegraphics[width=8cm]{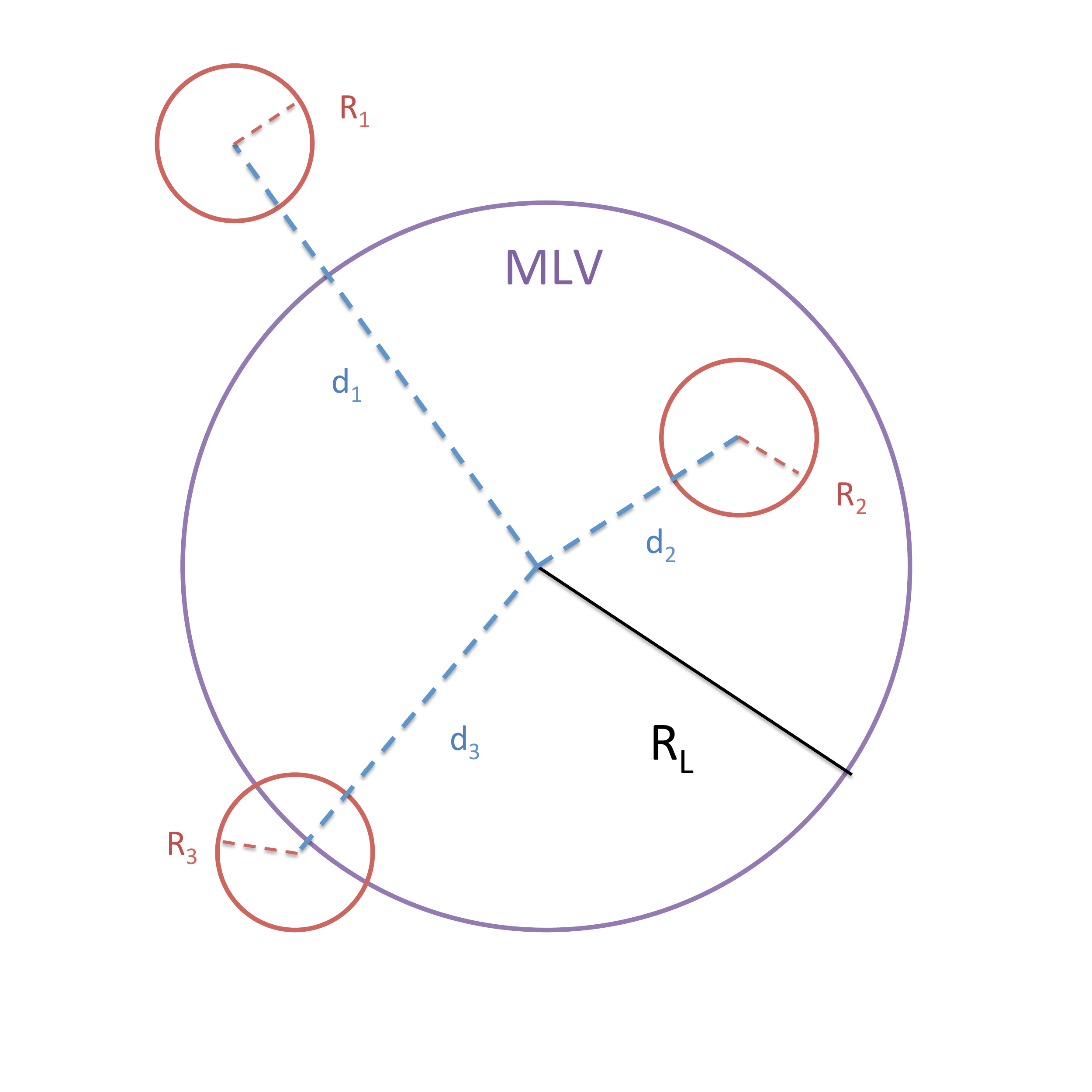}
  \caption{
    Illustrating the enrichment model for an atomic cooling
    halo (Section~\ref{secmix:ach}). The Mixed Lagrangian Volume (MLV) has a Lagrangian radius
    $R_L=6.4$\,kpc at $z=10$. Not shown is the size of the ACH
    Lagrangian Volume (ALV), which has a radius $\RLACH=11.4$\,kpc at $z=10$.
    Here, three minihaloes have experienced SN
    explosions. In this example, $q_1 = 0$, $q_2 = 1$, and $q_3$ is between
    0 and 1. The distances $d_i$ are uniformly drawn from a spherical
    volume $\mathcal{V}$ (not shown) whose radius is larger than
    $R_L+\rm{max}\{R_i\}$.
    \label{fig:spheremodel}}
\end{figure}

As seen in Section~\ref{sec:env}, most second-generation stars 
form out of gas that has been enriched by multiple SNe, even in the 
simplest possible environments of minihaloes (MH) and 
atomic cooling haloes (ACH).
Thus, there are two fundamental questions we must answer to understand
when a unique Pop~III chemical abundance signature can be preserved:
{\it (1)} How much does the initial mixing of multiple Pop~III SN ejecta
dilute this signature? And,
{\it (2)} how long does it take for Pop~II SNe to erase the Pop~III chemical signature?

We answer these questions with the following framework.
Consider two types of SNe, a normal type $n$ and a
special type $s$. Let $p_s$ be the probability that a given SN is
special. Then the yields from SN $i$ are given by:
\begin{equation}
\pmb{y}_i = 
\begin{cases}
  \pmb{y}_s & \text{probability}\ p_s \\
  \pmb{y}_n & \text{probability}\ 1-p_s, \\
\end{cases}
\end{equation}
where $\pmb{y}_i$ is the vector of metal yields for SN $i$ in units of
solar masses (\emph{not} the mass fraction). In this paper,
$\pmb{y}_n$ will always denote the metal yields from standard Pop~III
CCSNe (i.e. the first column of Tables~2 and 3). $\pmb{y}_s$ will
be the metal yields for faint SNe, wind+SNe, or PISNe.

To determine the fraction of stars preserving the special signature
during the initial enrichment period, we run these yields through an
enrichment model that determines the metallicity of second-generation
stars. We consider a mass $\Mmix$ within which the gas is chemically
homogeneous due to turbulent mixing. If the $j$th second-generation
star is enriched by $N_j$ Pop~III SNe, then the metallicity
$\pmb{Z}_j$ is given by
\begin{equation} \label{eq:metallicity}
  \pmb{Z}_j = \frac{1}{\Mmix} \sum_{i=0}^{N_j}q_i \pmb{y}_i,
\end{equation}
where $q_i \in [0,1]$ is the fraction of metals from the $i$th SN
that is mixed into $\Mmix$.
Depending on the environment, there will be different ways to
determine $\Mmix$, $N_j$, and $q_i$. We describe simple models
for these quantities for second-generation star formation in minihaloes
(Section~\ref{secmix:mh}) and atomic cooling haloes
(Section~\ref{secmix:ach}). 

To examine question {\it (1)}, we simplify the problem as follows. 
For a given metallicity $\pmb{Z}_j$, we classify the star as special
or normal depending on its metal ratios. When investigating the
CEMP signature, we consider the star to be special if $\CFe >
0.7$. When investigating the PISN signature, we consider the star to
be special if $\mbox{[Co/Ni]} < -0.5$.
For each value of $p_s$, we employ a Monte Carlo procedure to run our
enrichment model $N_{\rm total}$ times to produce
$\{\pmb{Z}_j\}_{j=1}^{N_{\rm total}}$. Each of these Monte Carlo
realizations represents a chemically homogeneous star cluster, of
which $N_s$ star clusters are classified as special. Then we define
the preservation fraction $f_s$ to be the fraction of these star
clusters that retain the special signature:
\begin{equation}
  f_s(p_s) = \frac{N_s}{N_{\rm total}},
\end{equation}
where we have made explicit the dependence of $f_s$ on the input
probability for special yields, $p_s$. The function $f_s(p_s)$ in different
environments thus describes how well a special signature is preserved,
depending on the likelihood for special SNe to occur.
In this paper we use $N_{\rm total}=10^5$ throughout.

To answer question {\it (2)}, in Section~\ref{secmix:popII}, we describe when
the contribution from Pop~II SNe is able to erase the unique Pop~III signature.

\subsection{Second-generation star formation in minihaloes} \label{secmix:mh}
All Pop~III SNe in a minihalo explode near the centre, driving a cumulative
blastwave. If fallback occurs and triggers second-generation star formation,
some fraction of the metals from each SN will enrich this second
generation \citep{Ritter12,Jeon14}.
When modeling the chemical abundances of metal poor stars,
the most common assumption is that metals from different SNe
stay completely unmixed, and abundance ratios reflect yields from a
single SN (e.g. \citealt{Joggerst10,Nomoto13}). This
idealized case implies $f_s(p_s) = p_s$. However, as we argued in
Section~\ref{sec:env}, this is likely not the generic case, even in
minihaloes. In general, $\sim 1-10$ SNe will enrich a minihalo. We
select $N_j$ from the distributions in Figure~\ref{fig:sfr2}.

To mix multiple SNe, the simplest assumption is that all metals from
all SNe completely mix into some gas mass
(e.g. \citealt{Tinsley80,Cooke14}). In the language of
Equation~\ref{eq:metallicity}, this fully-mixed model implies $q_i =
1$ for all SNe.

However, a more interesting and likely case is that not all metals
from each SN will fall back. 
\citet{Ritter15} investigate the scenario of recollapse in a minihalo
after seven SNe have exploded in the centre. About half the SN ejecta
escape from the halo, suggesting a similar fraction of the metal mass
escaping. The recollapsing gas in these simulations is biased towards
metals from the first SNe to explode in the minihalo, since this
ejecta runs into a denser surrounding medium and thus is able to cool
faster. The difference in fallback is a factor of $2-3$ between the
first and last star to explode.
The amount of metal fallback depends on the details of star formation,
stellar feedback, and hydrodynamics in each minihalo. 
This suggests that a stochastic model is required to describe the
amount of fallback. 
For simplicity, we model $q_i$ in the partial metal fallback case as a random
variable uniformly distributed between 0 and 1.
To approximate the effect of different SN energies,
we also consider a biased case where all the metals from faint SNe are
retained in the minihalo ($q_i=1$ if the $i$th SN is special), but the
normal SNe only retain a uniformly distributed fraction of their
metals ($q_i~\sim~\rm{Unif}[0,1]$ if the $i$th SN is normal).
A suite of more detailed calculations of metal mixing and fallback
such as those suggested in \citet{Ritter15} may find that other
fallback distributions are more realistic.

\begin{figure}
  \includegraphics[width=8cm]{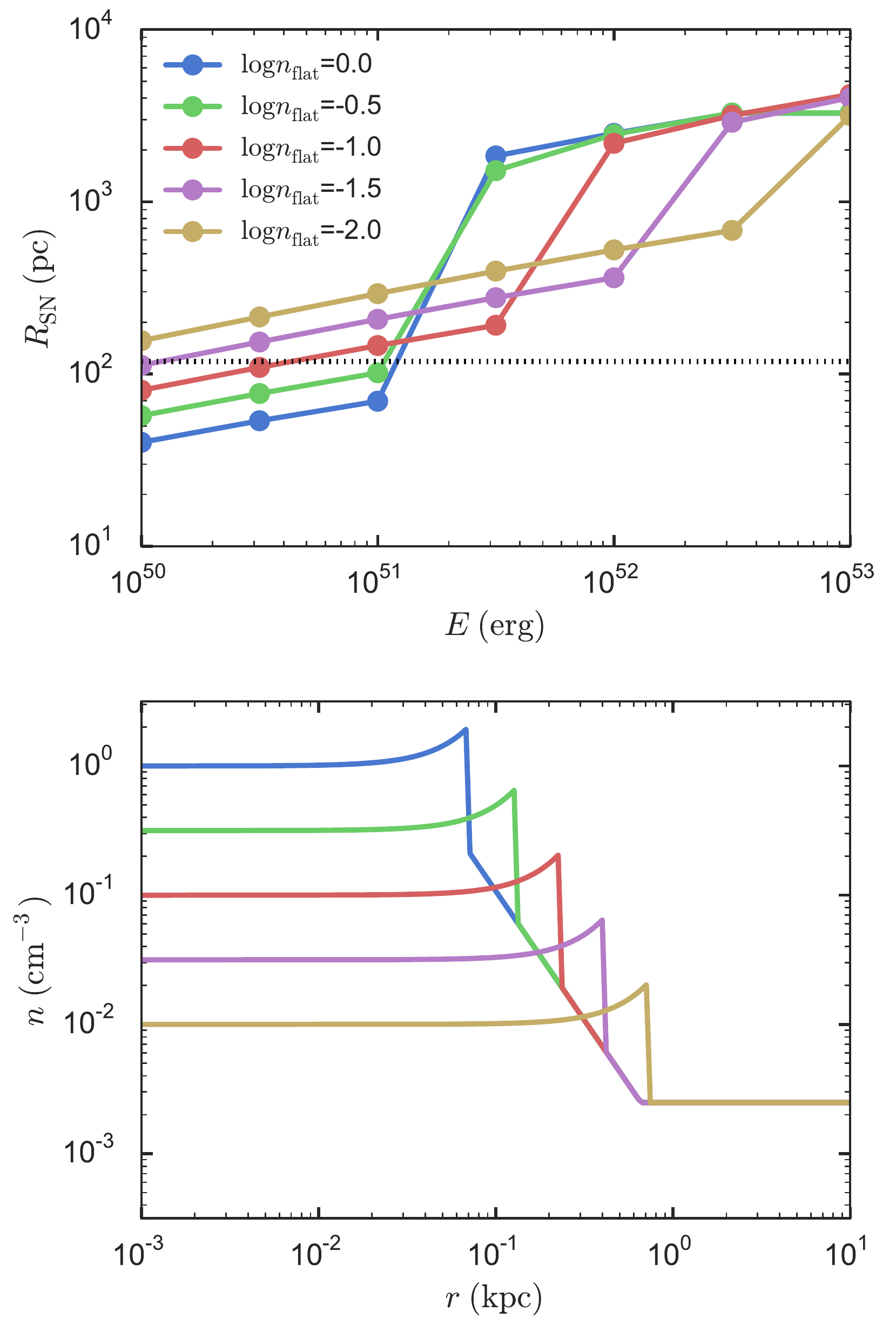}
  \caption{
    Top: SN bubble final radius expanding out of a minihalo as a
    function of initial energy deposited and ionizing feedback. The
    ionizing feedback is parametrized in terms of $\nflat$, which is
    the flat density in the inner part of the champagne solution.
    The dotted line indicates the virial radius of the minihalo.
    Bottom: density profiles for different $\nflat$. The temperature
    is $10^4$\,K throughout, which affects the gas pressure.
    \label{fig:SN_mix}}
\end{figure}

\subsection{Second-generation star formation in atomic cooling haloes}\label{secmix:ach}
Metal enrichment for second-generation star formation in an ACH is a
complex process that generally requires a full hydrodynamic simulation
(e.g. \citealt{Greif10,Wise12a,Jeon15}). However, given the
significant parameter uncertainties in Pop~III star formation, we here
describe a much simpler model that captures the physical nature of
this process. The model is illustrated in Figure~\ref{fig:spheremodel}.

Consider an ACH of virial mass $\MACH \sim 10^8 \Msun$ that virializes at
$\zvir=10$. If we trace all particles within the ACH to higher
redshift, there is an associated Lagrangian volume that encompasses all the
mass within the ACH at $\zvir$. We call this volume the \emph{ACH
Lagrangian Volume}, or ALV for short. This Lagrangian volume is
conceptually the same Lagrangian volume that is sampled at
high-resolution in a zoom-in simulation (e.g. \citealt{SafShrad14b}).
If the Lagrangian volume is spherical, there is some Lagrangian radius
$\RLACH$ such that
\begin{equation}
  \RLACH = \left( \frac{3 \MACH}{4\pi \rho_m} \right)^{1/3},
\end{equation}
where $\rho_m$ is the total matter density evaluated at some
redshift. 
The physical size of this Lagrangian radius at $\zvir$ is
found by using $\rho_m(\zvir)$, giving $\RLACH = 11.4$\,kpc.

Now consider the volume $\mathcal{V}$ surrounding and including the
ALV. We assume that star-forming minihaloes form uniformly distributed
within $\mathcal{V}$, the SNe in these minihaloes explode
independently, and the spherical SN remnant bubbles do not interact
with each other hydrodynamically.
Given the density of star-forming minihaloes $n$, the number of
minihaloes in the volume $\mathcal{V}$ is given by a Poisson
distribution with mean $n \mathcal{V}$. We draw $N_j$ from this
Poisson distribution.

Turbulence in the ACH will uniformly mix metals in much of the ACH
gas \citep{Greif10}, but not all the gas within the ACH is
well-mixed (i.e. $\Mmix \leq f_b \MACH$). We show in
Section~\ref{secmix:mixmass} that $\Mmix \sim 3 \times 10^6 \Msun$. The
Lagrangian volume corresponding to the mixed gas is correspondingly
smaller. We call this volume the \emph{Mixed Lagrangian Volume}, or
MLV. The radius of the MLV is denoted $R_L$, a smaller radius than $\RLACH$.

For a single minihalo, the fraction of its metal content that is mixed into
the MLV can be estimated by calculating the volume overlap between its
SN bubble and the Lagrangian volume of the MLV. In other words, if the 
$i$th SN bubble has radius $R_i$, the MLV has radius $R_L$, and the
centre of the minihalo is located at a distance $d_i$ from the centre of the MLV, then
\begin{equation}
  q_i = \frac{V_{\rm o}}{4\pi/3 R_i^3}
\end{equation}
where the overlap volume $V_{\rm o}$ can be found by simple geometry
and is given by
\begin{equation}
  V_{\rm o} = \begin{cases}
    0 & \text{ $d_i>R_{\rm sum}$} \\
    \frac{\pi (R_{\rm sum}-d_i)^2
      (d_i^2 + 2d_iR_{\rm sum}-3R_{\rm diff}^2)}{12d_i} &
    \text{ $R_{\rm diff} < d_i < R_{\rm sum}$} \\
    \frac{4 \pi}{3} R_i^3 & \text{ $d_i < R_{\rm diff}$}
    \end{cases},
\end{equation}
where $R_{\rm sum}=R_L+R_i$ and $R_{\rm diff}=|R_L-R_i|$.
For each minihalo, we draw $d_i$ from a uniform 3D spatial
distribution, where the maximum distance is given by taking
$\mathcal{V}$ as a sphere.
A uniform distribution is appropriate as the distance is considered
prior to the collapse of the ACH (see Figure~\ref{fig:spheremodel}).
Although the SN bubble radius expands with time, we do not consider
the time dependence and instead let the radius instantaneously expand to its
final (fixed) radius, only depending on the SN explosion energy. This is further
explained in Section~\ref{secmix:SN_rad}.

Other than $p_s$, the only parameters needed are $n$ and $\mathcal{V}$.
We choose $n$ so that $\bar{N}_{\rm MH}$, the mean number of minihaloes whose
centre is in the ALV, is 10, 30, and 100. This corresponds to $n
\approx 0.0016$, $0.0048$, and $0.016$ minihaloes per kpc$^3$ at $z=10$.
We choose $\mathcal{V}$ to be a sphere around the MLV whose
radius is 3.5 kpc larger than the MLV (8 kpc when we increase the SN
bubble radius beyond that). Since the number of SNe is Poisson, the
exact choice of $\mathcal{V}$ does not matter as long as it is large
enough to enclose $R_L+R_i$.

In practice, the mass of the SN bubble is likely concentrated
in a thin shell, and mixing into the spherical volume depends on the
development of Rayleigh--Taylor instabilities \citep{Madau01}. If no
such instabilities develop and all the metals are concentrated in the
shell, the overlap fraction is instead determined by the overlapping
solid angle, given by:
\begin{equation}
  q_i = \begin{cases}
    0 & \text{ $d_i>R_{\rm sum}$} \\
    \frac{2 R_i d_i - d_i^2 - R_i^2 + R_L^2}{4 R_i d_i} & \text{ $R_{\rm diff} < d_i < R_{\rm sum}$} \\
    1 & \text{ $d_i < R_{\rm diff}$}
    \end{cases}
\end{equation}
We have checked that using the thin-shell formula does not make a
significant difference in our results.

Since multiple SNe explode in each minihalo, for each SN bubble we
combine the yields from $\sim 1-10$ SNe with the 
$\rm{M}_*=100 \Msun$ number distribution in Figure~\ref{fig:sfr2}.

\begin{figure}
  \includegraphics[width=8cm]{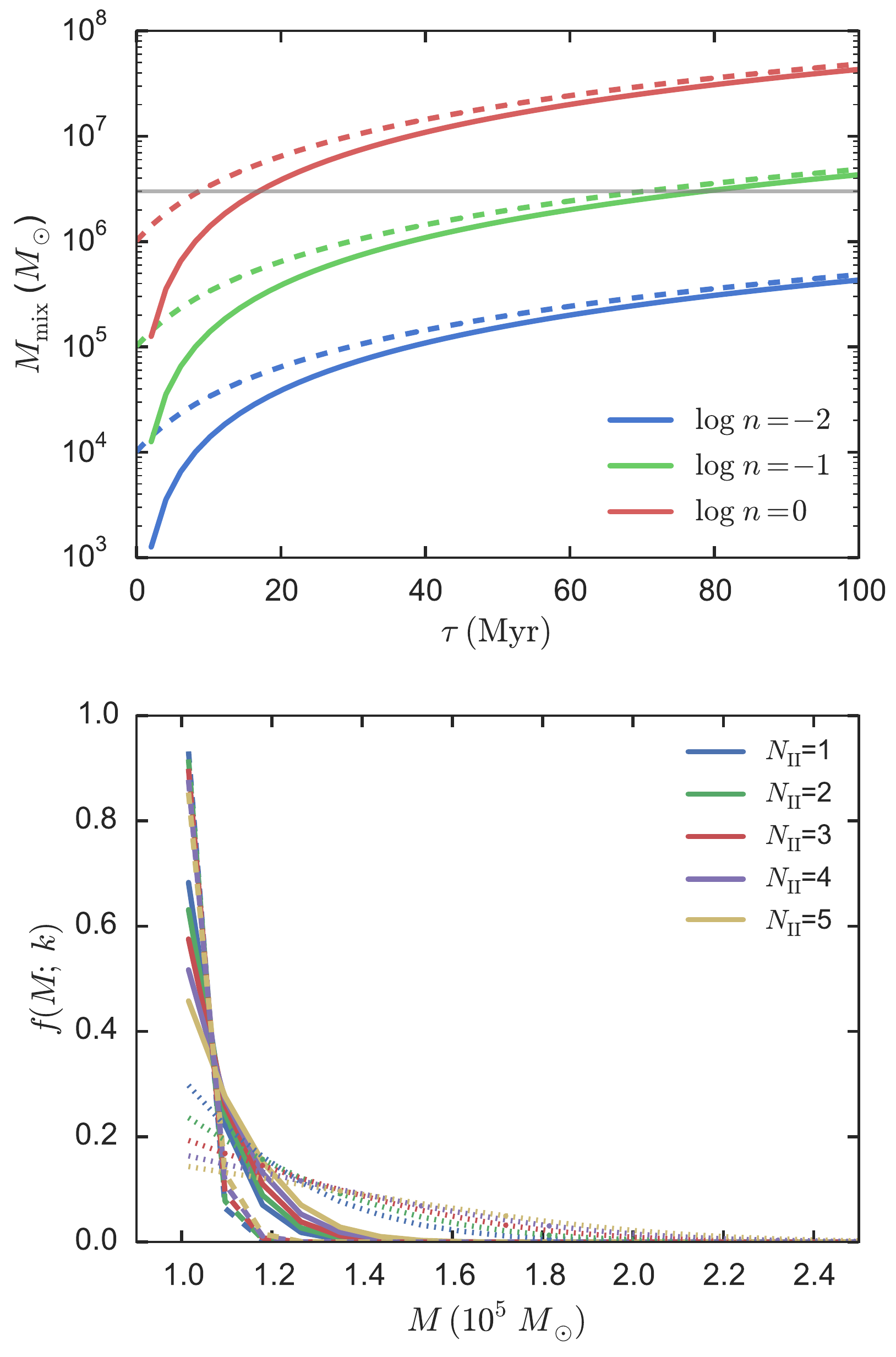}
  \caption{
    Top: mixing mass within an atomic cooling halo. 
    Solid lines indicate the mixing due to turbulent diffusion, which
    we use to determine the MLV.
    Light gray line indicates our choice for the MLV of $3 \times 10^6
    \Msun$.
    Dashed lines indicate how the mixing mass changes if including a SN
    bubble of 200 pc, which is relevant for calculating the
    \citet{Karlsson05} model.
    Bottom: mixing mass distributions for the three Pop~II star
    formation rates in the bottom panel of
    Figure~\ref{fig:sfr2}. Solid, dashed, and dotted lines correspond
    to models A, B, and C respectively. $k$ is the number of Pop~II
    SNe enriching a given low mass star. All distributions are roughly
    between $1-2 \times 10^5 \Msun$.
  \label{fig:Mmix}}
\end{figure}

\subsubsection{Supernova bubble radius} \label{secmix:SN_rad}
To calculate the size evolution of a Pop~III SN bubble, we use
the thin-shell approximation in an NFW halo \citep{Madau01}. We consider
haloes of mass $10^6 \Msun$ virializing at $z=25$.
The NFW concentration parameter is determined using
the model from \citet{Correa15} to be $2.6$, although note that
minihaloes are strongly triaxial and a spherical NFW density may not
properly characterize the gravitational potential \citep{Sasaki14}.
Different concentrations do not greatly affect the bubble radii.

To properly model the gas density profile, we must account for 
ionizing radiation from the Pop~III progenitor stars \citep{Bromm03b}.
To approximate the typical gas density profile resulting from
ionizing radiation, we use the self-similar champagne flow solution of
\citet{Shu02} as implemented by \citet{Wang12}. We parametrize the
self-similar scale using $\nflat$, the density in the inner
region. A larger $\nflat$ corresponds to a smaller stellar
lifetime \citep{Alvarez06}. Outside of the shock, we use an isothermal profile ($\rho 
\propto r^{-2}$), evaluated at the halo virial temperature.
When the isothermal halo density reaches the ambient IGM value, we
prevent it from decreasing further.
We use a gas temperature of $10^4$ K throughout, corresponding to
ionized gas since the HII region outpaces the SN blast wave.
We use the cooling function of
\citet{Sutherland93} for collisional ionisation equilibrium with $\FeH =-3.0$.
The bubble is initialized with a Sedov--Taylor solution at $10^4$~yr.
Unlike the original \citet{Madau01} model, we include only one
SN burst, so we do not include the time-dependent mechanical
luminosity term. Instead, SNe contribute their energy to the initial
energy of the SN bubble.
We take the stopping radius to be the point where the shock
velocity is equal to the sound speed in the photoionized medium
($T=10^4\,$K).
Although this model is simple, it reproduces the more detailed
calculations of \citet{Greif07}, where the SN bubble stalls at $2.5$\,kpc
for a $10^{52}$\,erg explosion in a $5 \times 10^5 \Msun$ minihalo.

The results are shown in Figure~\ref{fig:SN_mix}. 
The top panel shows the SN bubble radius as a function of total
energy in the bubble, while the bottom panel shows the gas density
profiles that the bubble pushes against. Not shown is the underlying
NFW profile, which slows the bubble gravitationally.
If a bubble is energetic enough to break past the champagne flow
shock, it is able to travel much further into the IGM before
stopping.
In all cases with $E \leq 10^{52.5}$\,erg, the bubbles reach their maximum
radius in less than $100$\,Myr.
Since the time delay between Pop~III star formation and virialization
of the ACH is $\sim100$\,Myr (see Figure~\ref{fig:sfr1}), this
justifies using an instantaneous bubble size in our ACH enrichment model.
Based on Figure~\ref{fig:SN_mix}, a good fiducial radius for a
$10^{51}$\,erg CCSN explosion is $\sim 0.2-0.4$\,kpc, while a $10^{52}$\,erg PISN
explosion reaches $\sim 3$\,kpc. We use $0.2$\,kpc and $3$\,kpc as the
inital radius for CCSNe and PISNe, respectively.
However, if the SN bubble expands far enough from the minihalo to be
swept up by the ambient Hubble expansion, the radius can grow by a
factor of up to $2.5$.
The maximum extent of the SN bubbles will then be $1$\,kpc and
$7.5$\,kpc for CCSNe and PISNe, respectively. We show models
using these larger radii as well.

\subsubsection{Mixing mass} \label{secmix:mixmass}
The size of the Mixed Lagrangian Volume in an atomic cooling halo is 
primarily determined by the extent of turbulent diffusion \citep{Greif10}.
To model this, we use the mixing volume formula from
\citet{Karlsson08}
\begin{equation} \label{eq:Vmix}
  V_{\rm mix}(\tau) = \frac{4\pi}{3}\left(6 D_t \tau\right)^{3/2}
\end{equation}
where $\tau$ is the time over which turbulent diffusion occurs, and
$D_t$ is the turbulent diffusion coefficient.
The factor of $6$ is a combination of a factor of 2 from the 1D
solution to the diffusion equation and a factor of 3 from three
spatial dimensions \citep{Karlsson13}.
The turbulent diffusion coefficient is given as 
\begin{equation}
  D_t \sim \langle\vturb\rangle \lturb/3
\end{equation}
where $\vturb$ is a typical turbulent driving velocity, and $\lturb$ is a
typical turbulent driving scale. 
This is an order-of-magnitude estimate based on dimensional grounds,
similar in nature to that used by mixing length theory
to describe convective transport in stars where the mixing length is
parametrized in terms of a characteristic scale height (e.g.
\citealt{Kippenhahn12}).
Since turbulence is driven by gravity in our haloes
\citep{Wise07,Greif08,SafShrad12,Ritter15}, we estimate that the
characteristic $\vturb$ is given by the halo virial velocity, and the
characteristic $\lturb$ is given by a fraction of the virial radius
$\sim R_{\rm vir}/10$. Values for the diffusion coefficient are given in
Table~\ref{tbl:env}.
The mixing mass as a function of $\tau$ and the ambient gas density
are shown in Figure~\ref{fig:Mmix}.

A typical gas density in an atomic cooling halo is
$n \sim 10^{-1}$\,cm$^{-3}$, and a typical amount of time for the gas
to be turbulently mixed is the halo free fall time, $t_{ff} \approx
0.1/H(z) \approx 70$\,Myr for $z=10$.
This results in a typical mixing mass of $\Mmix = 3 \times 10^6 \Msun$ of
gas, or $R_L = 6.4$\,kpc.

Note that there is in principle a lower bound to the size of the MLV,
since a single star cluster is chemically homogenous 
(e.g. \citealt{BlandHaw10,Feng14}). In the Milky Way this is
emprically found to be $\sim 10^5 \Msun$, but it could be lower in the
earliest galaxies \citep{SafShrad14b}.

\subsection{Contamination by Pop~II supernovae} \label{secmix:popII}
Since Pop~II CCSNe and standard Pop~III CCSNe produce similar
metal yields, it may be difficult to distinguish true second
generation stars from low mass stars that have been contaminated by
Pop~II SNe.
This can happen in massive minihaloes as well as in atomic cooling haloes,
where potential wells are sufficiently deep to retain their gas if SN feedback
is not too strong. At this point, the halo is massive enough to constitute a
self-enriching system, a characteristic of a first galaxy \citep{Bromm11}.
To answer question {\it (2)}, how quickly Pop~II SN can erase unique
Pop~III signatures, we add Pop~II SNe one at a time until the
signature is wiped out.
Here, as before, it is important to know how much gas the SN ejecta
mixes into.

For a massive minihalo, in our spherically symmetric model a single
$10^{51}$ SN fills the entire virial radius. Thus we can instantly
mix the Pop~II SNe into the entire minihalo volume.
For simplicity, we assume the total gas mass available is the same as
in the original minihalo, although in principle gas will be continually accreted.

In an atomic cooling halo, the mixing mass for an individual SN is
often assumed to be the maximum radius of a supernova
remnant (e.g. \citealt{Ryan96}). However, turbulent mixing can expand
the mixing mass. To estimate the size of this effect, we use the 
\citet{Karlsson05} model to calculate mixing mass distributions using
the three Pop~II  star formation models in Figure~\ref{fig:sfr2}.
We employ the mixing volume formula from
\citet{Karlsson08}, which is similar to Equation~\ref{eq:Vmix} but now
accounts for the inital SN bubble size $R_{\rm SN}$:
\begin{equation} \label{eq:VmixRSN}
  V_{\rm mix}(\tau) = \frac{4\pi}{3}\left(6 D_t \tau  + R_{\rm SN}^2\right)^{3/2}.
\end{equation}
We use $R_{\rm SN}=200$ pc from Figure~\ref{fig:SN_mix}, and $n =
0.12$\,cm$^{-3}$ and $D_t = 7 \times 10^{-4}$\,kpc$^2$\,Myr$^{-1}$ from
\citet{Karlsson05}. 
The results for a small number $N_{\rm II}$ of Pop
II SNe are plotted in Figure~\ref{fig:Mmix}. Increasing $N_{\rm II}$ allows for slightly
higher mixing masses, but the typical mixing masses
are $1-2 \times 10^5 \Msun$, only slightly larger than 
the mixing mass right after the SN bubble stops expanding. 
Lowering the star formation rate by a factor of 10 only slightly
increases the maximum mixing mass here to $\sim 3 \times 10^5 \Msun$.
Thus, it is reasonable to assume that the mixing volume for Pop~II SNe is
about the mass of its SN remnant, which we take to be $3 \times 10^5
\Msun$ to maximize the dilution. It is only at larger $N_{\rm II}$
that the effect of turbulent mixing will be important.

\begin{figure*}
  \includegraphics[width=16cm]{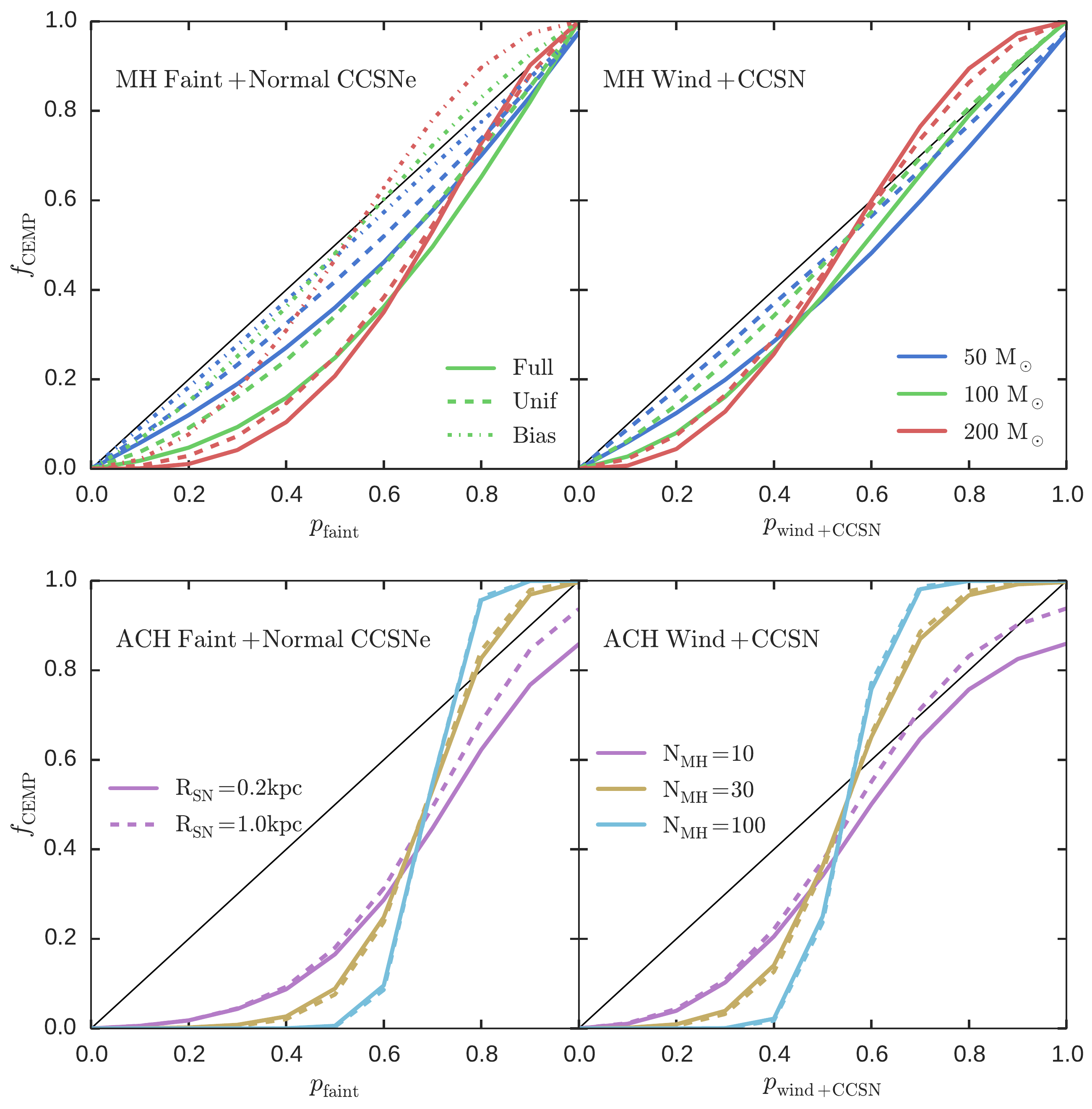}
  \caption{CEMP signature preservation in star-forming haloes after
    the addition of metals by Pop III SNe. Top row: Minihalo models
    are shown. The solid, dashed, and dot-dashed lines indicate
    different amounts of SN ejecta fallback into the;
    minihalo. Different colours indicate different total mass of
    Pop~III stars formed per minihalo, which translates into the SN
    number distributions in Figure~\ref{fig:sfr2}. Overall, the
    observed $f_{\rm CEMP}$ in second-generation stars reflects the
    input special fractions $p_s$. Bottom row: atomic cooling halo
    models are shown. The solid and dashed lines indicate different SN
    bubble sizes. Different colors indicate different values for the
    mean number of minihaloes contributing metals to an ACH. In both
    cases, $f_{\rm CEMP}$ rapidly changes from 0 to 1 over a narrow
    range of input $p_s$. See text for more discussion.
    \label{fig:CEMPpopIII}}
\end{figure*}

\begin{figure*}
  \includegraphics[width=16cm]{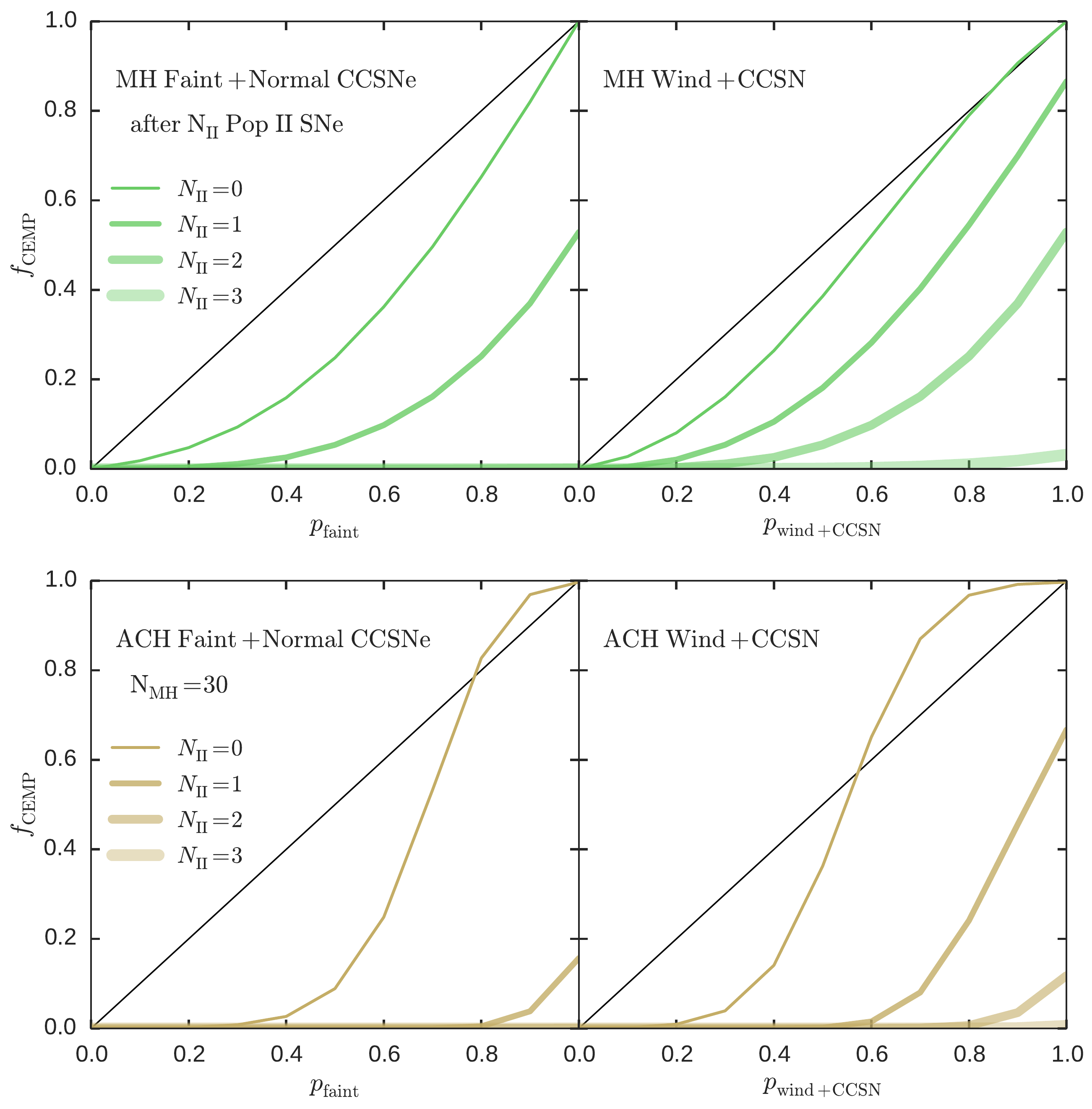}
  \caption{CEMP signature preservation after dilution by $N_{\rm II}$
    Pop~II SNe.
    Panels are arranged as in Figure~\ref{fig:CEMPpopIII}. 
    Different line thickness and transparency denote different numbers
    of Pop~II SNe.
    The MH lines show deviations from the model using $\rm{M}_*=100
    \Msun$ of Pop~III stars per minihalo and full metal fallback in
    the minihalo after adding Pop~II SNe.
    The ACH models show deviations from the model using 
    $R_{\rm SN}=200$\,pc and $\bar{N}_{\rm MH}=30$.
    In all models, only a few Pop~II SNe are required to eliminate the
    CEMP signature, even if 100 per cent of Pop~III stars produce the CEMP
    signature. Note that in several panels, the larger $N_{\rm
      II}$ lines are at $f_{\rm CEMP}\approx 0$ regardless of $p_s$ and thus are
    barely visible at the bottom of the panels.
    \label{fig:CEMPdiluted}}
\end{figure*}

We can also come to the conclusion that turbulent mixing is relatively
unimportant by estimating the time-scale for turbulent mixing to
dominate the mixing volume in Equation~\ref{eq:VmixRSN}. This is
approximately when
\begin{equation}
  \tau = \frac{R_{\rm SN}^2}{6 D_t} \approx
  27\text{\,Myr}
  \,\left(\frac{R_{\rm SN}}{200\,\text{pc}}\right)^2
  \,\left(\frac{D_t}{10^{-3}\,\text{kpc}^2\text{Myr}^{-1}}\right)^{-1}
\end{equation}
This time-scale is longer or comparable to the $\sim 10$\,Myr recovery time in a
self-enriching system, so turbulent mixing will not affect the mixing
mass by more than a factor of a few.

\section{Results} \label{sec:results}
 We now present model calculations showing how well the CEMP
(Section~\ref{secres:CEMP}) and PISN (Section~\ref{secres:PISN})
chemical signatures are preserved in gas from which metal-poor stars would
form in the early Universe. We inject a natal fraction $p_s$ of
special Pop~III yields along with $1-p_s$ of normal Pop~III CCSN. We process
this through our mixing models (see Section~\ref{sec:mix}), and calculate
the fraction $f_s$ of stars retaining the special Pop~III signature.
We consider the preservation both during the initial enrichment by
combining multiple Pop~III SNe, as well as after adding
Pop~II SNe. In Section~\ref{secres:obsCEMP}, we compare our models
from Section~\ref{secres:CEMP} to the observed stellar halo CEMP star
fraction.

\subsection{Preserving a CEMP signature} \label{secres:CEMP}
As a reminder, we say a metal-poor star displays the \emph{CEMP
signature} if it has $\CFe > 0.7$. The input special fraction $p_s$ will refer to two different
models for producing the CEMP signature in Pop~III stars. If $p_s$ is
$p_{\rm faint}$, then we combine faint and normal CCSNe to produce the
CEMP signature. If $p_s$ is $p_{\rm wind+CCSN}$, then we add a carbon wind to
the yields of some of the CCSNe to produce the CEMP signature.

\subsubsection{Multiple Pop III SNe}
In Figure~\ref{fig:CEMPpopIII}, we show the preservation fraction for
the CEMP signature in the MH and ACH environments (top and bottom
rows, respectively) assuming the signature is generated through either
the faint+normal CCSN or the wind+CCSN mechanisms (left and right
columns, respectively). As an example of interpreting the figure,
consider the model of faint+normal CCSNe in a MH (top-left
corner). Suppose $p_{\rm faint}=0.6$, i.e. 60 per cent of \emph{all} Pop~III
SNe are faint CCSNe and 40 per cent are normal CCSNe. Then the solid green
line shows that if minihaloes form a combined $100 \Msun$ of Pop~III
stars and all the metals fully mix ($q_i=1$), then $\sim 35$ per cent of
metal-poor stars forming in minihaloes will retain the CEMP signature.
For reference, the thin black line in the plots indicates where $f_{\rm
  CEMP} = p_s$, or where the CEMP signature has been perfectly
preserved. We stress that all the second-generation stars in any
given minihalo will have the same signature, since they form from a
chemically homogeneous star cluster \citep{Ritter15,SafShrad15}.

In the top row of Figure~\ref{fig:CEMPpopIII} we show the CEMP signature
preservation fraction for the minihalo models. In general, the MH
Wind+CCSN models are within $\sim 10$ per cent of perfect preservation.
However, the faint+normal CCSNe models can deviate strongly from
perfect preservation. In particular, at low $p_{\rm faint}$ and larger
numbers of Pop~III SNe, the CEMP signature is strongly
suppressed. This is simply because it is easier to reduce the $\CFe$
ratio by adding iron from the normal CCSNe to the iron-poor faint CCSNe.
As expected, having biased fallback for faint SNe (dot-dashed lines)
results in much better signature preservation.

In the bottom row, we show the CEMP signature preservation fraction
for the atomic cooling halo models. The solid and dashed lines show SN
expansion radii of $200$\,pc and $1$\,kpc, showing the range of
preservation fractions based on whether the Hubble flow expands the SN
bubbles. The colors indicate different numbers of minihaloes that
contributed metals to the ACH. Recall that each minihalo in these
models can produce multiple CCSNe, where the number is drawn from the
green $100\Msun$ distribution in the top panel of
Figure~\ref{fig:sfr2}. While the MH models reasonably preserve the
Pop~III signature regardless of $p_s$, the ACH models show a sharp
transition from poorly preserved to over-preserved. This is due to the
average $\CFe$ ratio in all haloes being determined by $p_{\rm
  faint}$. As we increase the number of SNe contributing metals to a
gas reservoir (MH or ACH), the abundance scatter between different gas
reservoirs decreases, resulting in a larger fraction of the gas reservoirs being
close to the average value. Note that in the $\bar{N}_{\rm MH}=10$
case, when $p_{\rm faint}$ and $p_{\rm wind+CCSN} = 1$ there is still
not perfect preservation. This occurs because of cases where all
minihaloes are within the ALV but do not intersect the MLV.

Overall, it is generally the case that the CEMP signature can be
preserved in the gas after combining yields from multiple Pop~III SNe
in MHs. It is difficult but not impossible to preserve the
signature in ACHs. However, even if second-generation low-mass
stars can be identified from observations of metal-poor stars, our
models show that the observed CEMP star fraction will generally be
slightly lower than the actual fraction of Pop~III SNe producing the
CEMP signature. A metal enrichment model (such as the one used to
calculate Figure~\ref{fig:CEMPpopIII}) is required to convert
observed CEMP star fractions into the actual fraction of Pop~III SNe
producing the CEMP signature.

\begin{figure}
  \includegraphics[width=9cm]{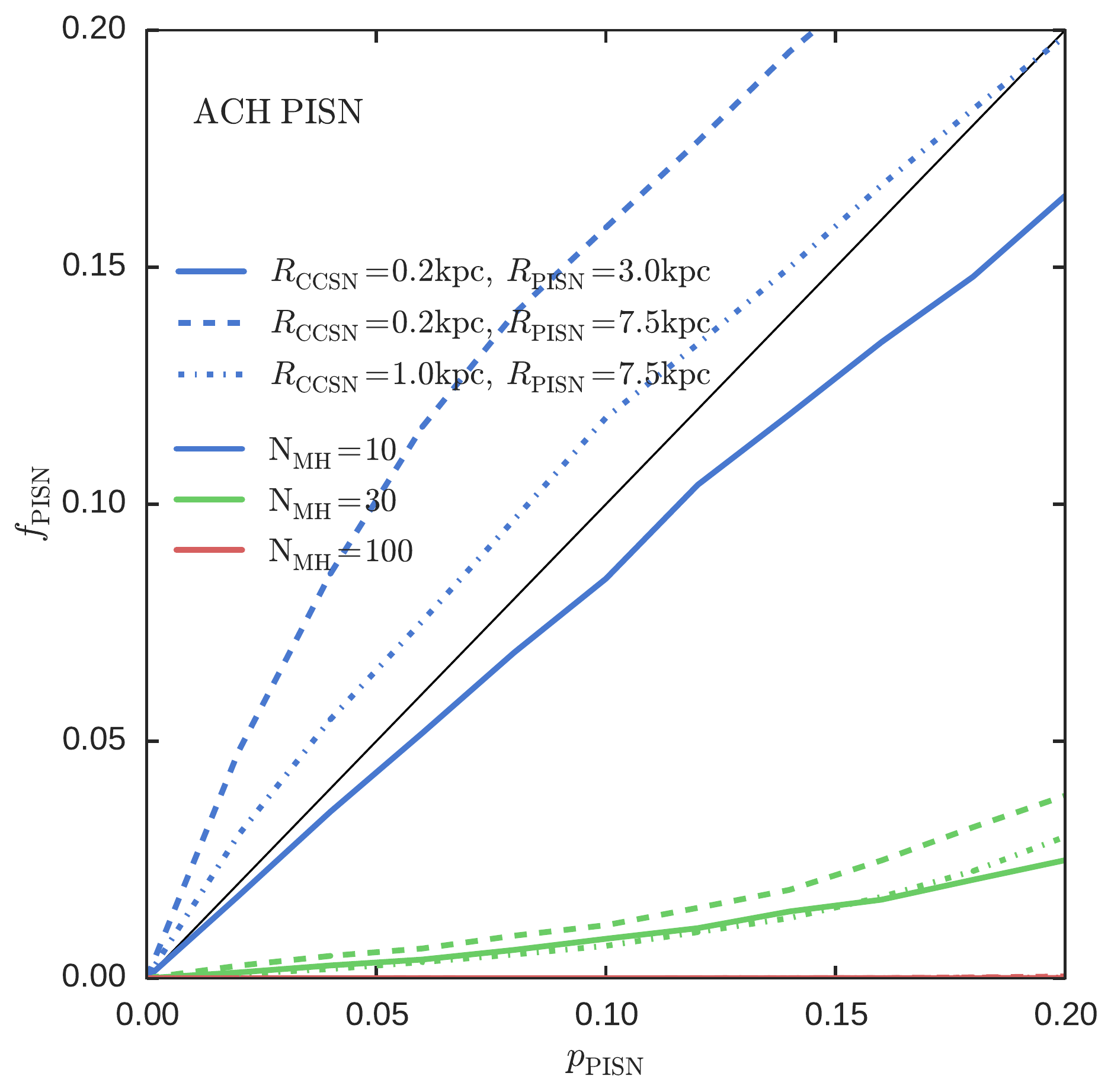} \\
  \includegraphics[width=9cm]{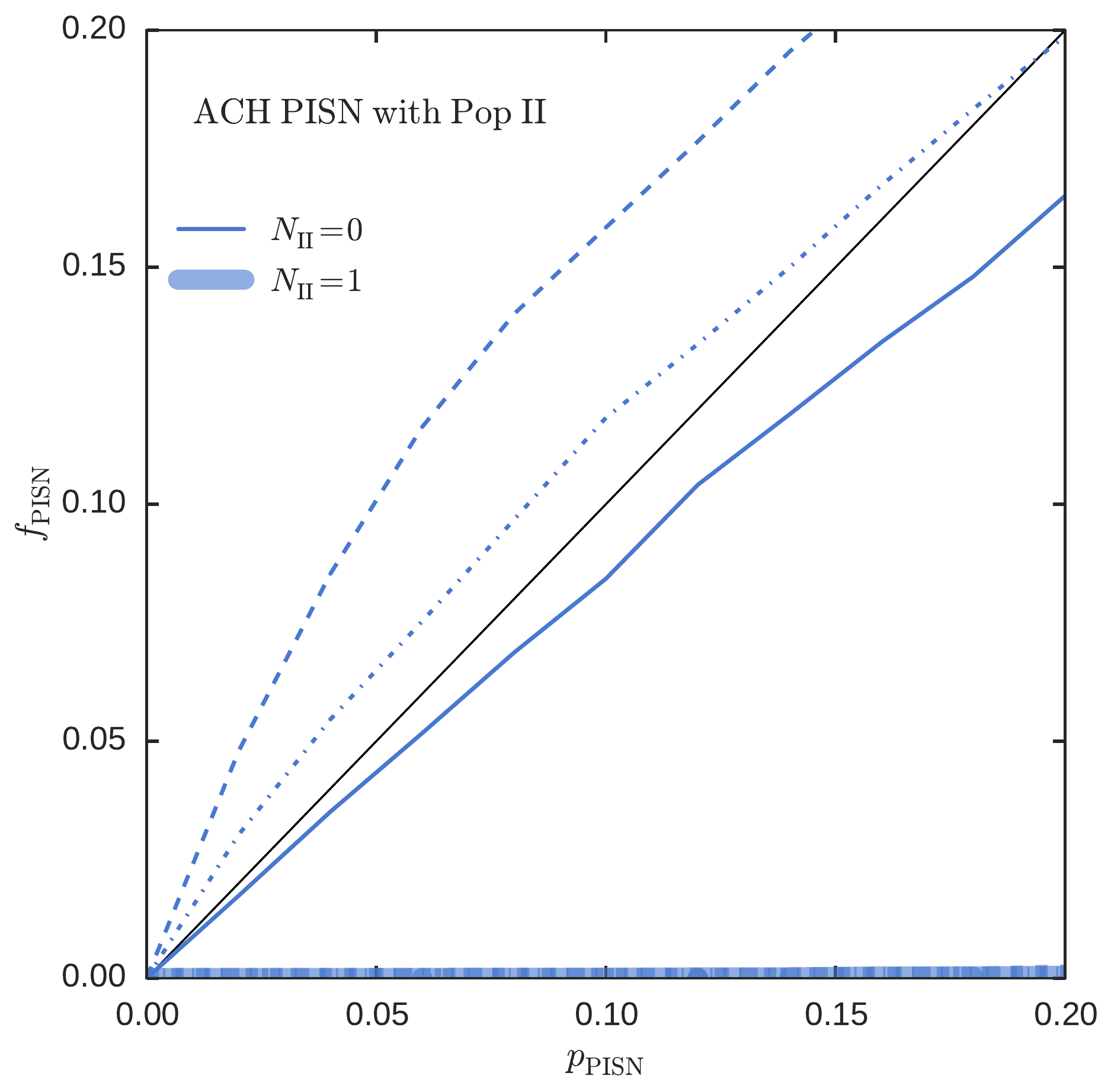}
  \caption{
    PISN signature preservation in the ACH model. 
    Top: Pop~III only.
    The line colors indicate different values for the mean number of
    minihaloes. Solid, dashed, and dot-dashed lines indicate different
    SN bubble radii.
    The thin black line denotes perfect preservation.
    If a minihalo exeriences a PISN, that is the only SN in the minihalo. If it has CCSNe, then
    the number of CCSNe is drawn from the $100 \Msun$ distribution of
    Figure~\ref{fig:sfr2}.
    Note that $p_{\rm PISN}$ only extends from 0.0--0.2, as we do not
    expect most minihaloes to host PISNe.
    Bottom: after adding a single Pop~II SN, the PISN signature is
    erased. The $N_{\rm II}=1$ line has $f_{\rm PISN} \approx 0$ for all shown
    $p_{\rm PISN}$ and is thus barely visible at the bottom of the plot.
  \label{fig:MLVpisn}}
\end{figure}

\begin{figure*}
  \includegraphics[width=16cm]{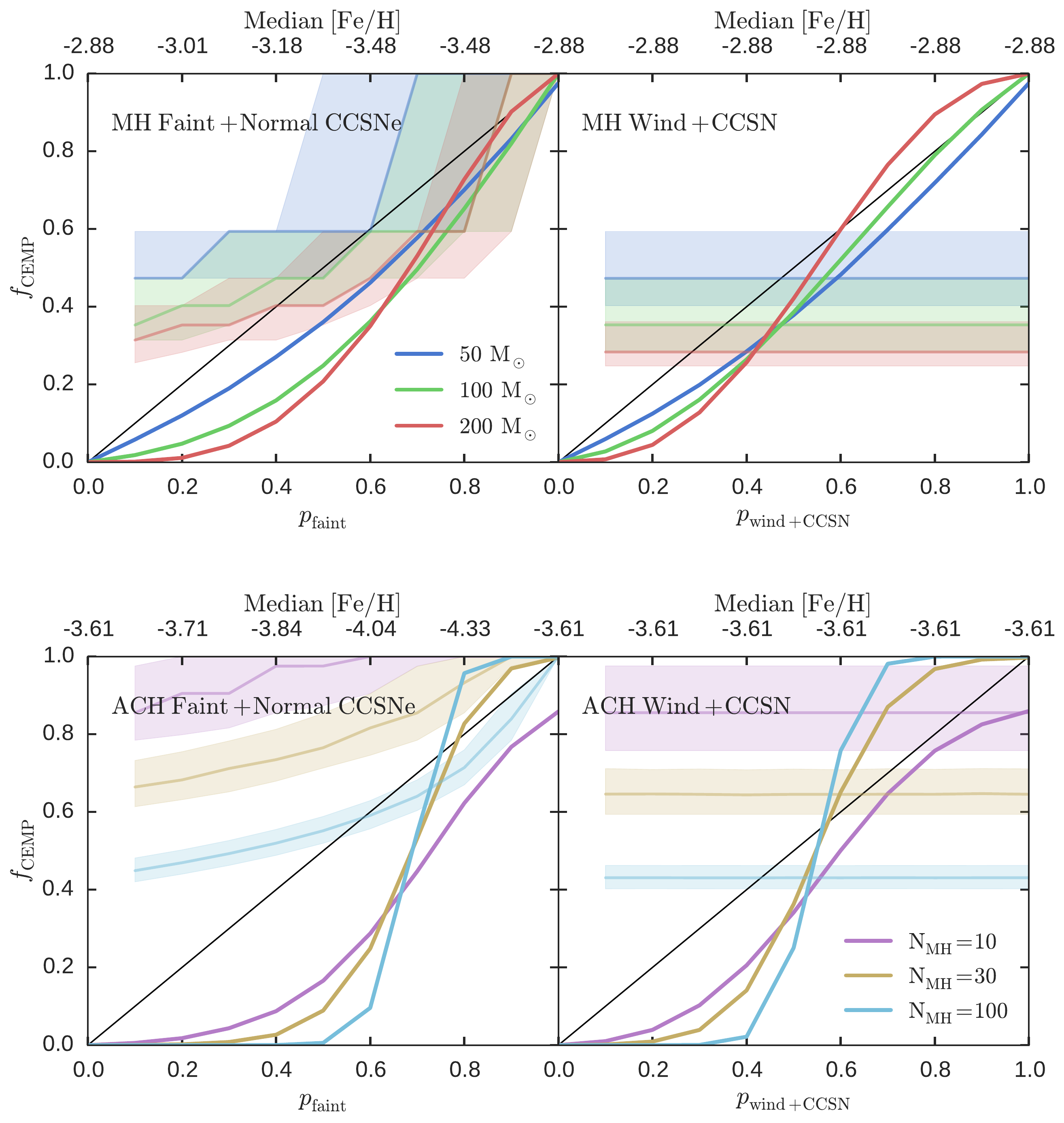}
  \caption{CEMP signature preservation compared to expected CEMP star
    fractions based on model $\FeH$ and observed CEMP star fractions
    (Equation~\ref{eq:obscfrac}).
    Shaded regions show the 25 and 75 percentile of the model $\FeH$
    distributions translated into observed CEMP star fractions, with light solid
    line showing the median value.
    The median $\FeH$ for the $100 \Msun$ and $\bar{N}_{\rm MH}=30$ models
    are shown on the top axis of each panel.
    Panels are arranged as in Figure~\ref{fig:CEMPpopIII}. For the MH
    panels, we only show the full fallback case. For the ACH panels,
    we only show $R_{\rm SN}=0.2$\,kpc.
    The values of $p_{\rm faint}$ and $p_{\rm wind+CCSN}$ where the
    model lines intersect the shaded bars are consistent with the
    observed CEMP fraction.
    \label{fig:CEMPobs}}
\end{figure*}

\subsubsection{Adding Pop II SNe}
In Figure~\ref{fig:CEMPdiluted}, we show how the preservation
signature is kept after adding Pop~II SNe yields to the metal distributions
from Figure~\ref{fig:CEMPpopIII}.
Each Pop~II SN yield is diluted into $3 \times 10^5 \Msun$, which we
have conservatively chosen to be on the high end of possible dilution masses.
Interestingly, in all four cases, adding $1-3$ Pop~II SNe is enough to erase the
CEMP signature in the gas as imprinted by Pop~III SNe.
In the MH case, this is because the signature is already close
to being erased by a few normal Pop~III CCSNe.
In the ACH case, it is a result of the $\sim 10$ times discrepancy in the
mixing mass between Pop~III and Pop~II SNe.

A typical Pop~II burst in these environments produces $\sim1000 \Msun$
of stars. For a Salpeter IMF, there are $\sim 5$ massive stars that
explode as SNe in the first Pop~II star formation
burst. As a consequence, this strongly suggests that if CEMP stars
trace Pop~III nucleosynthesis, they are truly second-generation stars
that form out of only Pop~III metals (although multiple Pop~III stars
will contribute metals).

\subsection{Preserving a PISN signature}\label{secres:PISN}
We define a metal-poor star as retaining the PISN signature if it
forms from gas with $\mbox{[Co/Ni]} < -0.5$. As previously
discussed, this is a representative odd-even element abundance ratio.
In the top panel of Figure~\ref{fig:MLVpisn}, we show the preservation
fraction for the PISN signature ($f_{\rm PISN}$) depending on the
fraction of minihaloes hosting PISNe ($p_{\rm PISN}$). If a minihalo
does not host a PISN, it then has a random number of CCSNe drawn from
the $100 \Msun$ distribution of Figure~\ref{fig:sfr2}.
Models are shown for different SN bubble radii and different numbers
of minihaloes. We only use the ACH model and not the MH model, since
the strong PISN explosion evacuates its minihalo of gas \citep{Whalen08}.

In general, the PISN signature is well preserved for the
expected number of $10$ minihaloes contributing to the ACH. 
As more minihaloes are added, the number of normal Pop~III
CCSNe increases and is able to overwhelm the PISN signature at these
low $p_{\rm PISN}$ fractions. Note that when the size of the PISN
bubbles is increased, the PISN signature is actually stronger than
perfect preservation. This is because for the same number density of
minihaloes (as parametrized by $\rm{\bar{N}_{\rm MH}}$), increasing the SN
bubble radius allows more such bubbles to overlap with the MLV. The MLV
is then able to sample more PISN bubbles than CCSN bubbles. Increasing
the size of the CCSN bubbles somewhat counteracts the PISN signature
for the same reasons. 

The PISN signature is very fragile. The bottom
panel of Figure~\ref{fig:MLVpisn} shows that a single Pop~II SN is
able to wipe out the PISN signature. Again, this is due to the vastly
different dilution factors between the Pop~III and Pop~II SNe.

 It is interesting to compare the results of this model to cosmological
 simulations, such as the one by \citet{Jeon15}. In contrast with
 many previous simulations, this simulation allows for both PISNe and
 CCSNe to occur in the minihaloes. By $z \approx 10.5$, 2 PISNe have formed
 out of a total of $\sim 200$ Pop~III stars, corresponding to $p_{\rm PISN} \approx
 0.01$. These two PISNe contribute the majority of metals in the
 simulation, but nearly all of these metals are ejected into the IGM
 rather than found in virialized haloes. Their most massive galaxy is
 thus predominantly enriched by CCSNe. We also find that the
 $\rm{\bar{N}_{\rm MH}}=30$ case (similar to their number of minihaloes) has
 CCSNe dominating the metal enrichment in a typical ACH when $p_{\rm
   PISN} = 0.01$.

\subsection{Interpreting observations of metal-poor stars} \label{secres:obsCEMP}

 Ultimately, the goal of stellar archaeology is to interpret abundances
of metal-poor stars to learn about properties of Pop~III
stars. We now present an example of how our results in
Section~\ref{secres:CEMP} can be used to interpret observed CEMP star
fractions in the Milky Way halo as constraints on Pop~III properties.

Let us suppose that the CEMP star fraction in the stellar halo is
representative of second-generation stars. In practice, the halo is an
amalgam of stars from many sources, and even at fairly low
metallicities some halo stars may form after the first few generations. 
Regardless, with this assumption, we can compare the observed CEMP star
fraction to that obtained by our metal enrichment models. Our 
models give estimates of the $\FeH$ distributions of Pop~II stars.
We take these $\FeH$ distributions and find the 25th and 75th
percentile values to define a range of $f_{\rm CEMP}$.
We consider models to be allowed if their predicted $f_{\rm CEMP}$ is
consistent with the $f_{\rm CEMP}$ computed from the model $\FeH$
distributions.
For the observed CEMP star fraction,
we take an approximate fit to the CEMP fraction from
figure~16 of \citet{Placco14} for $\CFe > 0.7$.
\begin{equation} \label{eq:obscfrac}
  f_{\rm CEMP} = \begin{cases}
    0.2 & \FeH > -2.5 \\
    -0.8 - 0.4 \times \FeH & -4.5 < \FeH < -2.5 \\
    1.0 & \FeH < -4.5
  \end{cases}
\end{equation}

In Figure~\ref{fig:CEMPobs} we show the preservation fraction for some
of the models in Figure~\ref{fig:CEMPpopIII}.
The shaded bars in the background indicate the
observed CEMP star fraction evaluated at the modelled $\FeH$ distributions.
Note that the shaded bars for the wind+CCSN models are flat, since
every SN has the same iron yield.

When the model lines overlap the shaded bars, the corresponding
$p_{\rm faint}$ or $p_{\rm wind+CCSN}$ are consistent with the
observations. Taking the figure at face value, we see that the MH
models produce the expected carbon fractions, although it generally
requires a fairly high $p_s$. The faint+normal CCSNe models require
$p_{\rm faint} \gtrsim 0.6$. If this picture is correct, then our
results imply that $\gtrsim 60$ per cent of Pop~III stars exploded as a faint
SN. The high fraction of faint SNe can be interpreted as an indirect
tracer of the IMF (e.g. \citealt{Heger02}). A similar calculation for
the biased fallback model suggests that $p_{\rm faint} \gtrsim 0.5$.
If instead the CEMP signature is produced in MHs by carbon winds, the
wind+CCSN models require $0.7 \gtrsim p_{\rm wind+CCSN} \gtrsim
0.4$. This could be interpreted as the fraction of Pop~III stars that
are rotating rapidly enough to have significant winds.

In contrast, the ACH models seem to be unable to produce
consistent CEMP star fractions without a huge number of minihaloes
contributing to the ACH. Recall that each minihalo here can produce
multiple Pop~III SNe, and $\rm{\bar{N}_{\rm MH}}=100$ is a strong upper limit
assuming there has been no smooth mass accretion.
Thus, if the CEMP signature is created solely by Pop~III stars, a
large fraction of Pop~III stars must produce the high $\CFe$
abundances, and most of these stars form in minihaloes.

Given the simplicity of our enrichment models, we caution that 
more sophisticated models could somewhat affect the exact numbers.
In particular, while the overall metallicity in our
models is likely correct to within an order of magnitude, the actual value
can be shifted by changing the dilution mass within some reasonable
range. From Equation \ref{eq:obscfrac}, a 0.2 dex change in the model
$\FeH$ results in a 0.08 change in the observed $f_{\rm CEMP}$.
It is also a fairly strong assumption that the observed CEMP stars
represent only second-generation stars, although at
$\FeH \lesssim -3$ it is probably reasonably accurate. However, we can
qualitatively see the effect of including subsequent generations of
stars. Since the CEMP signature is rapidly erased by Pop~II SNe, any
contaminating stars will reduce the observed CEMP star fraction. Thus a
true second-generation CEMP star fraction will be higher than the shaded
bars plotted in Figure~\ref{fig:CEMPobs}.
This further increases the $p_s$ required to be consistent with the MH
models, as well as exacerbating the inability of the ACH models to
produce the observed CEMP star fractions.
Regardless, this approach of comparing observed special fractions
$f_s$ to those found from models is an emerging framework to test where our
understanding needs to most improve to constrain properties of Pop~III
stars.

\section{Summary and Conclusions} \label{sec:conclusion}
In this study we focused on whether unique Pop~III chemical signatures
can be preserved in the gas from which the first low mass stars form.
Using simple models that capture the essence of initial Pop~III
chemical enrichment, we calculate the fraction of second-generation
stars preserving unique Pop~III chemical signatures depending on the
input fraction of Pop~III stars that produce said signature.
We now summarize our main findings and conclusions.

\textbf{The chemical signature of Pop~III can be preserved in gas
  from which the first low-mass stars formed.}  Metal-poor stars are
believed to trace the nucleosynthetic yields of their progenitor stars
which in the case of the most metal-poor ones are individual Pop~III
SNe events. While this is the fundamental assumption that stellar
archaeology rests on, detailed investigations to what extent SN yield
signatures remain intact in the gas after the explosion and factoring
in various environmental aspects such as mixing and the presence of
multiple SNe are still lacking 

Even the very first enrichment step has irreducible
complexity, with at least a handful of Pop~III SNe contributing metals
to most second-generation star forming environments. We find that the
mixing and re-accretion of gas onto the halo following these SNe does
not always wipe out the original nucleosynthetic signatures, and
thus that it is possible for second-generation stars to trace
Pop~III properties. This validates the endeavor of stellar archaeology.
Nevertheless, it becomes increasingly clear that
interpreting observed abundance signatures in the most metal-poor
stars requires using chemical enrichment models to factor in effects
such as metal mixing and turbulence (see also
\citealt{Ritter15}). Otherwise the properties of Pop~III stars may not
be correctly recovered.

\textbf{The frequency of metal-poor stars with a special abundance
  signature (e.g. high [C/Fe]) cannot be directly interpreted as
  the fraction of Pop~III stars producing that signature.}  By
comparing the results of our MH and ACH models to the observed
fraction of CEMP stars in the Galactic stellar halo, we find that CEMP
stars likely formed from gas in a minihalo enriched by a SN that
underwent metal fallback. The observed high CEMP star fraction is not
necessarily produced by faint SNe alone, but if so then the fraction
of Pop~III stars producing the CEMP signature must be rather high, at
least $\sim 40-60$ per cent even though the observed CEMP star fractions are
lower. This high fraction could be an indication of a
biased scenario, such as preferential second-generation star formation
in minihaloes that experience low energy SNe \citep{Cooke14}. 
However, a better understanding of the CCSN explosion
mechanism is likely necessary to confirm this, as even quite energetic
SNe can produce a similar abundance pattern \citep{Tominaga07}.
Either way, it is nearly impossible to reproduce the observed CEMP
star fraction in atomic cooling haloes.

\textbf{Any Pop~III chemical signatures are quickly erased after the
  emergence of Pop~II SNe.} This fundamental result suggests that
stars with unique and unusual chemical abundance
signatures are clean probes of Pop~III SN yields. This effect has
been understood before in the general context of chemical evolution
(e.g. \citealt{Audouze95}), but the sheer rapidity with which unique
abundance signatures are erased has not been appreciated. Future
abundance studies based on high-resolution spectroscopy should focus
on any unusual halo stars as they will likely reflect Pop~III
nucleosynthesis yields. 
Moreover, stars with $\mbox{[Fe/H]}\lesssim-4.0$ of which many
show unique abundances can all be regarded second-generation
stars. This is in line with other suggestions from SN yield fitting of
observed abundance signatures in the most metal-poor stars
\citep{Tominaga14,Keller14,Placco15}, although these authors also
considered stars with halo-like abundances and not just unusual
signatures.

\textbf{A PISN signature is quickly wiped out by the emergence of
  Pop~II SNe.} A strong odd-even effect is believed to be a tell-tale
sign of a PISN signature. So far, only one star has been identified
that displays some degree of an odd-even effect \citep{Aoki14}. This
may suggest that PISNe are exceedingly rare. However, this does not
necessarily imply that Pop~III {\it PISN explosions} were equally rare, or
even absent. Our results instead suggest that any PISN enrichment pattern would
rapidly be hidden under subsequent layers of Pop~II core-collapse enrichment.
In addition, a PISN
signature might only occur in stars with high metallicity since the Ca
yield of a PISN is rather large \citep{Karlsson08}. Current selection
biases on the basis of a weak Ca\,II\,K line would thus preclude
finding these stars. 
Interestingly, even
though PISNe produce large
amounts of metals, it appears to be very difficult for most of the
PISN metals to be retained in a first galaxy environment that would
then lead to the formation of metal-poor stars. This is due to metals
being dispersed over a much larger volume compared to those from
CCSNe. As previously suggested, the best place to search for the PISN
signature may be in IGM gas at high redshifts with absorption
spectroscopy \citep{Jeon15}.

\vspace{0.3cm}

Despite the fact that our enrichment models are idealized and only
capture the essence of the problem, our results are encouraging and
highly relevant to the interpretation of the abundance signatures of
metal-poor stars and thus to the field of stellar archaeology. 
Hydrodynamic simulations that elucidate the nature of metal mixing and
enrichment in the early universe are particularly important. The
results from these simulations will allow for more quantitative
investigations since our simple models have already demonstrated that
one can recover valuable information about Pop~III stars from stellar
archaeology.

We have shown that it is possible to learn about the nature and
properties of Pop~III stars by using chemical abundance measurements
of metal-poor halo stars, such as in the case of the observed CEMP
star fraction. However, the complex formation history of the stellar
halo makes it difficult to isolate true second-generation stars. The
only way to best choose candidates is to select the lowest metallicity
stars, but that may not yield a complete sample since not all
second-generation stars may be at the lowest metallicities (see
e.g. \citealt{Frebel12,Frebel14}).

A more promising way to find and study second-generation stars may be
in dwarf galaxies and to carry out dwarf galaxy archaeology
\citep{Frebel12}. As each dwarf galaxy is a single, confined object
with one specific history, it will be easier to model the galaxies and to
separate second-generation stars from other stars within these
galaxies. Of the Milky Way's dwarf galaxies, the ultra-faint dwarfs
may be the best places to clearly identify second-generation
stars. Their chemical abundances and star formation histories suggest
that ultra-faint dwarfs form stars only over a very short period of
time \citep{Frebel14,Brown14}, and due to their inefficient star
formation, a large fraction of their stars (or even all of them) may
be second-generation. If ultra-faint dwarf galaxies were the 
descendents of a single Pop~III star-forming minihalo, each
ultra-faint dwarf galaxy would only preserve one single combination of
Pop~III yields. However, if these galaxies are ACHs that form from a
merger of more than one second-generation star forming minihalo, then
multiple second-generation gas reservoirs may be represented in the
dwarf galaxy.

A disadvantage of using stars in dwarf galaxies to constrain
properties of Pop~III stars is the relatively small number of stars
with available chemical abundance observations. High
resolution spectroscopy is the gold standard, but the faintness of
most dwarf galaxy stars may require further development of medium
resolution techniques (e.g. \citealt{Vargas13,Vargas14,Kirby15}) or
await the era of extremely large telescopes.

All but the least-luminous ultra-faint dwarfs have experienced some
level of extended star formation \citep{Vargas13}. Thus, for detailed
studies of true Pop~III signatures, it will be important to separate
bona-fide second-generation stars from subsequent star formation.
A promising approach may be to combine different Pop~III yields and enrichment
models with chemical evolution models that account for hierarchical
galaxy formation (e.g. \citealt{Salvadori15}). These models will need
to account for the effects that recent hydrodynamic simulations have
highlighted \citep{Ritter15,Sluder15,BlandHaw15}. With a larger sample
of available dwarf galaxy abundances and more sophisticated modeling
of the formation of these systems, it may become possible to robustly
and comprehensively constrain the properties of Pop~III stars with
dwarf galaxy archaeology.

\section*{Acknowledgements}
APJ thanks Ralf Klessen and Brendan Griffen for
helpful conversations, and the UC HiPACC 2014 Summer School for useful
lectures. This work has made extensive use of the python libraries
\texttt{numpy}, \texttt{scipy}, \texttt{matplotlib}, and
\texttt{seaborn}. APJ and AF are supported by NSF grant AST-1255160.
AF acknowledges support from the Silverman (1968) Family Career Development Professorship.
VB is supported by NSF grant AST-1413501. 

\bibliographystyle{mnras}

\label{lastpage}
\end{document}